\documentclass[9pt,twocolumn,twoside]{osajnl}

\journal{ao} 

\setboolean{shortarticle}{false}

\usepackage{graphicx}
\usepackage{indentfirst}
\usepackage{algorithm}
\usepackage{algpseudocode}
\usepackage{lineno}
\usepackage{subcaption}

\title{Morphological variations to ptychographic algorithm}

\author[1]{F. Salinas}
\author[1,*]{M. A. Sol\'is-Prosser}

\affil[1]{Departamento de Ciencias F\'isicas, Facultad de Ingenier\'ia y Ciencias, Universidad de La Frontera,\\
Avenida Francisco Salazar 01145, Temuco, Chile.}

\affil[*]{Corresponding author: miguel.solis@ufrontera.cl}

\begin{abstract}
Ptychography is a technique widely used in microscopy for achieving high-resolution imaging. This method relies on computational processing of images gathered from diffraction patterns produced by several partial illuminations of a sample. In this work, we numerically studied the effect of using different shapes for illuminating the aforementioned sample: convex shapes, such as circles and regular polygons, and unconnected shapes that resemble a QR code. Our results suggest that the use of unconnected shapes seems to outperform convex shapes in terms of convergence and, in some cases, accuracy. \\

\end{abstract}

\setboolean{displaycopyright}{false}

\begin{document}

\maketitle

{\bf See published version at} \href{https://doi.org/10.1364/AO.462173}{\bf https://doi.org/10.1364/AO.462173}

\section{Introduction}
Every time one tries to recover optical information about an object, one may think in taking a high-resolution picture, which will be able to inform us about some features of the aforementioned object: color and transparency. However, the phase is a nontrivial piece of information missing from a single picture, which is very important in some fields such as optical microscopy~\cite{zernike1942,shaked2013}, electron microscopy~\cite{rodenburg1989, rempfer1992, coene1992, opdebeeck1996, allen2000, allen2001a} and X-ray imaging~\cite{millane1990, fitzgerald2000, taylor2003, lewis2004, wu2005, burvall2011, wu2022}. In this context, the field of \emph{phase retrieval} aims to obtain the phase of a complex-valued function that describes either a wave field or the transmission function of an object. One of the first approaches, if not the first, was introduced by Gerchberg and Saxton in 1972~\cite{gerchberg1972}, which requires a picture of the object and a picture of its Fourier transform. This algorithm, also known as the GS algorithm, is guaranteed to converge, although very slowly, and it is not free of inaccuracies. Based on the GS algorithm, Fienup proposed an algorithm that only needs information of the diffraction pattern of the object~\cite{fienup1978,fienup1982}. These algorithms have been mathematically analysed in terms of convex optimization~\cite{bauschke2002} and, very recently, Zhao and Chi~\cite{zhao2020} introduced modifications to the GS algorithm that improved its convergence and accuracy, and studied their feasibility for optical cryptography.   

Other widely used technique is ptychography, which is a phase retrieval method that allows us to retrieve both amplitude and phase of a sample object function using data from several diffraction patterns, each obtained by illuminating a subregion of the object being reconstructed~\cite{hoppe1982,faulkner2004,rodenburg2004} and applying an iterative numerical algorithm on the obtained images, which is known as Ptychographical Iterative Engine (PIE)~\cite{faulkner2005}. For this method to work properly, it is crucial that the different regions illuminated in the sample have a significant overlap between each other \cite{rodenburg2008,bunk2008}. As the GS algorithm can reconstruct images--although with some drawbacks--from only two images, one may see that a dataset used for ptychography exhibits redundancy whenever more than two diffraction patterns are recorded. This redundancy, far from being undesired, allows one to achieve superresolved imaging~\cite{rodenburg2007,maiden2011}.

Ptychography has benefited from several improvements and modifications, including--but not limited to--enhanced algorithms such as extended PIE (ePIE)~\cite{maiden2009,maiden2012}, combination with a Hybrid Input-Output approach~\cite{konijnenberg2016}, a reciprocal approach in which the illumination beam is tilted instead of displaced on a sample, also known as Fourier ptychography~\cite{zheng2013, yeh2015, zhanglei-lei2017, konda2020, bianco2021, lee2021}, among others~\cite{nashed2014,odstrcil2016,maiden2017,yao2021}. This technique and its variants have already found application in the context of optical imaging~\cite{thibault2009,maiden2010, claus2013, zhang2013, godden2014,li2019,chang2020}, X-ray microscopy~\cite{rodenburg2007,rodenburg2007a,thibault2008, dierolf2010,dierolf2010a,edo2013,maiden2013,stockmar2015, morrison2018,pfeiffer2018,holler2019,kahnt2021,yao2021}, electron microscopy~\cite{nellist1998,rodenburg2007,haigh2009, hue2010, hurst2010, humphry2012,oleary2021}, optical encryption~\cite{shi2013,rawat2015,zhu2019}, and recent demonstrations show promising applications in Quantum Information Science~\cite{aidukas2019,fernandes2019,fernandes2020}.

Noteworthily, the partial illumination of the sample is circularly shaped in most works of the literature. One of the few works that studied other possibilities is the one of Ref.~\cite{wangya-li2013}, which included hexagonal and square shapes. The work of Ref.~\cite{huang2014} explored the impact of overlap uniformness in the quality of the reconstruction. Besides from these studies, and up to our knowledge, irregularly-shaped partial illuminations have not yet been explored. Moreover, as most iterative algorithms, convergence and accuracy might depend on the choice of an initial guess. In this work, we explored the effects of considering different shapes for the partial illuminations of the sample on both accuracy and convergence when the PIE algorithm is used. Through simulations, we considered squares, regular hexagons, circles, and irregular unconnected regions resembling a QR code. As a strategy to avoid reaching to misleading conclusions, we performed every reconstruction with 50 different initial guesses in order to have statistically significant results. Consequently, our results showed that continuous regions (polygons and circles) exhibit significant differences in performances only in a handful of cases. Unexpectedly, the use of unconnected regions, in general, outperformed the use of continuous regions.

This article is organized as follows. Section~\ref{sec:method} explains the method in detail. Particularly,  Subsection~\ref{sec:objects} introduced the images being used as the optical object to be reconstructed and the different shapes of the illumination functions; Subsection~\ref{sec:reconst} introduces useful notation for this article and explains how the initial guessed function are dealt with; Subsection~\ref{sec:FOM} shows an overview of the algorithm used and the figures of merit used to assess performance. Section~\ref{sec:results} shows the results of our study. Section~\ref{sec:conclusion} concludes the paper.

\section{Method}
\label{sec:method}
\subsection{Illumination functions and optical objects \label{sec:objects}}

\par
\begin{figure}[!ht]
	\centering
	\includegraphics[width=0.45\textwidth]{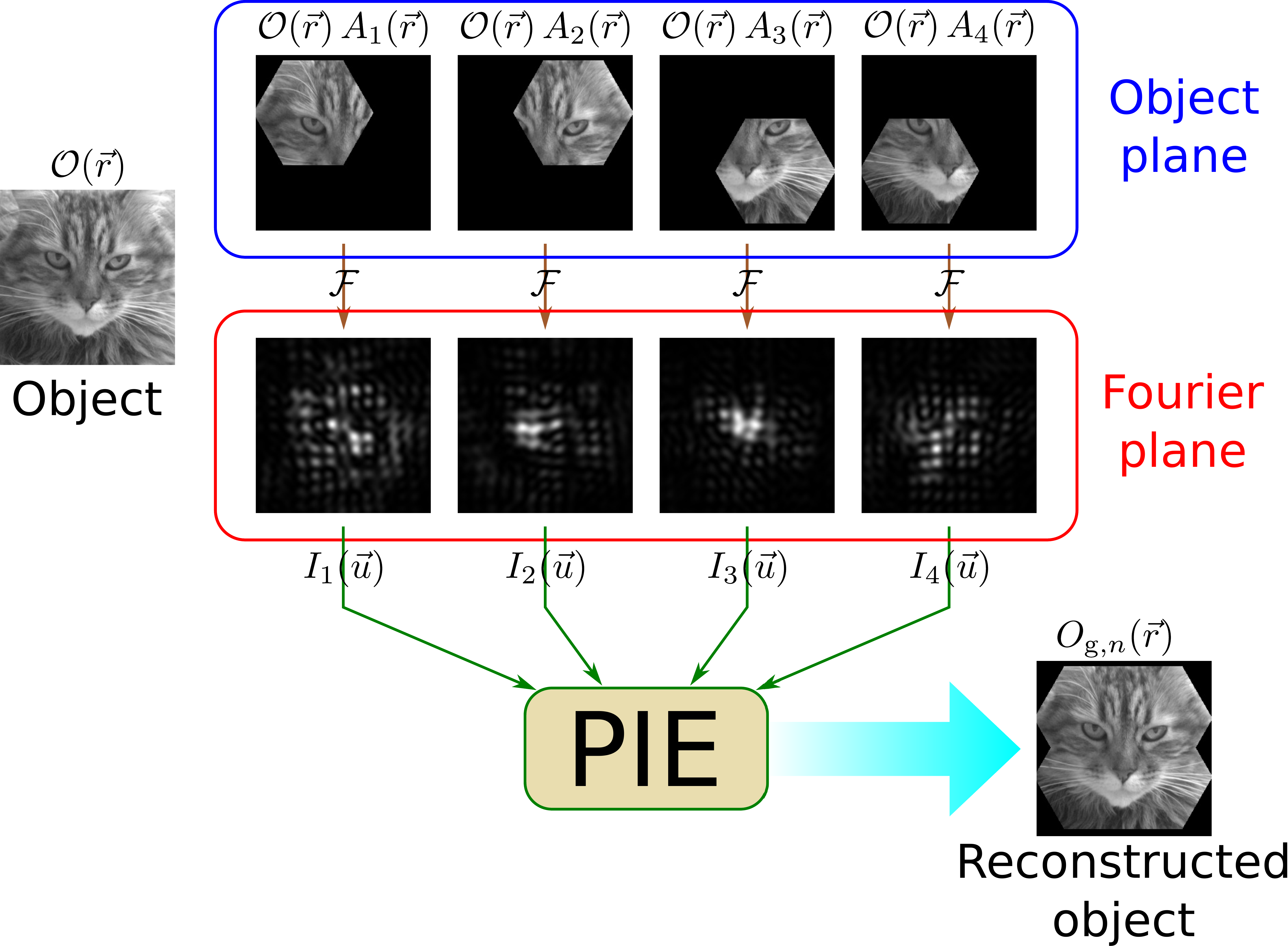}\\[5mm]
	\includegraphics[width=0.45\textwidth]{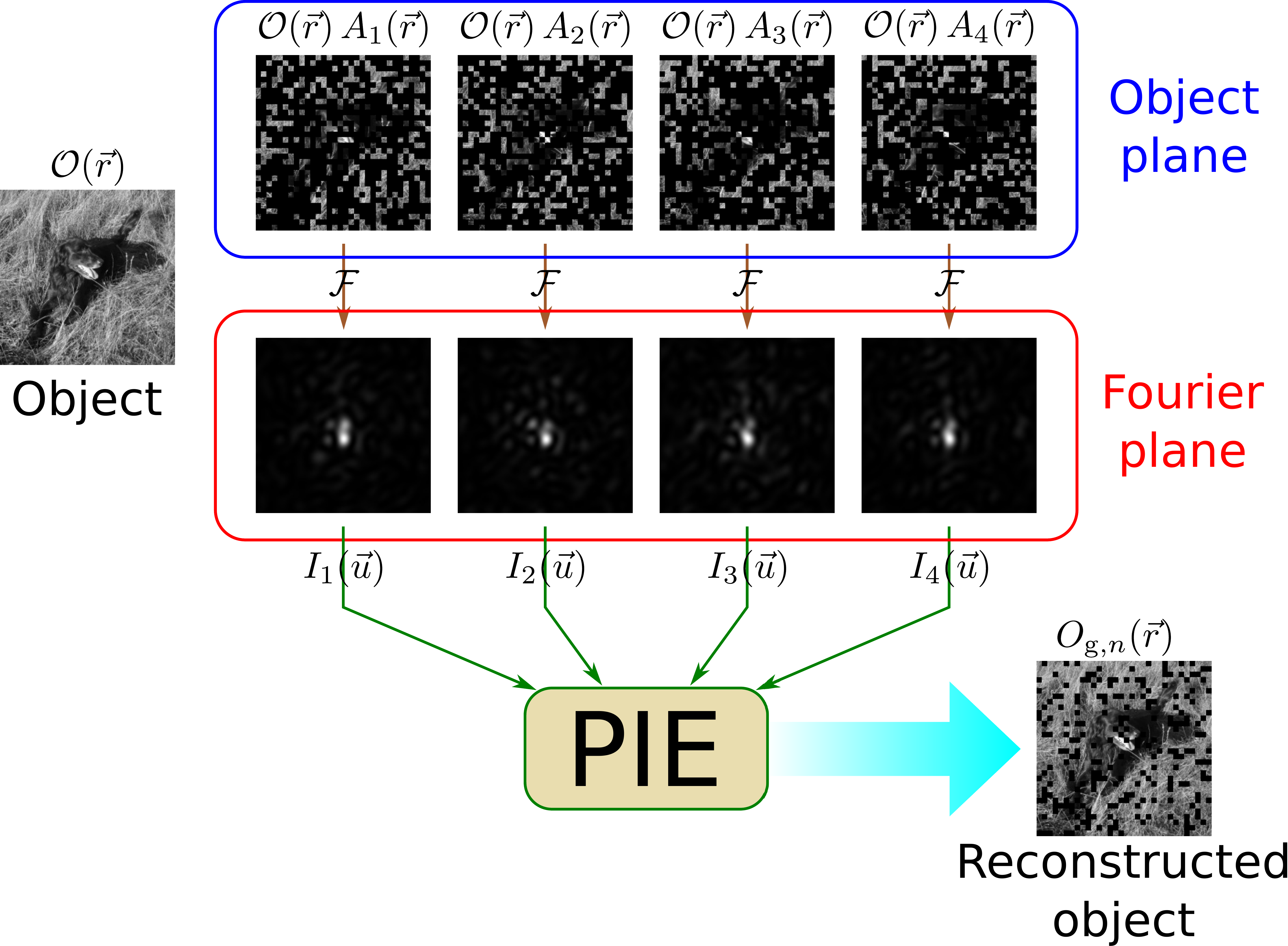}
	\caption{Simplified depiction of a ptychographic scheme. Upper half: convex regular shapes (4 hexagons in this example) are being used as illumination functions. The radius of each hexagon is such that its area is the closest possible to one of a circle of $R=80$ pixels as radius. Lower half: unconnected shapes (4 different ones in this example) are being used as illumination functions. For this particular example, $\ell=8$ pixels. The number of squares is such that the illuminated area in each $A_j(\vec{r})$ is the closest possible to one of a circle of $R=80$ pixels as radius.}
	\label{fig:Fig1}
\end{figure}

 Let $\mathcal{O}(\vec{r})$ be a 2D-transmission function of an arbitrary optical object (sample). For this work, we will also assume this object will be illuminated by a coherent monochromatic plane-wave electromagnetic field. This light field can be modulated through diffractive devices. In this context, let us define a set of illumination functions $\left\{A_j(\vec{r})\right\}_{j=1}^{N}$ which will describe the incident light field being shaped in order to illuminate different parts of the sample. In this context, we have tested two classes of illumination functions: (a) regular convex figures, and (b) unconnected sets. On one hand, in (a) we used $N$ illumination functions shaped as a regular figure, distributed among $N$ positions on the sample (see Figure~\ref{fig:Fig1}, left, for an example with $N=4$). For this purpose, we compared circular, hexagonal and square shapes. Let $R$ be the radius of the circles. Two values of $R$ were used in this study: 40 and 80 pixels.

\begin{figure}[!ht]
    \centering
    \includegraphics[width=0.75\columnwidth]{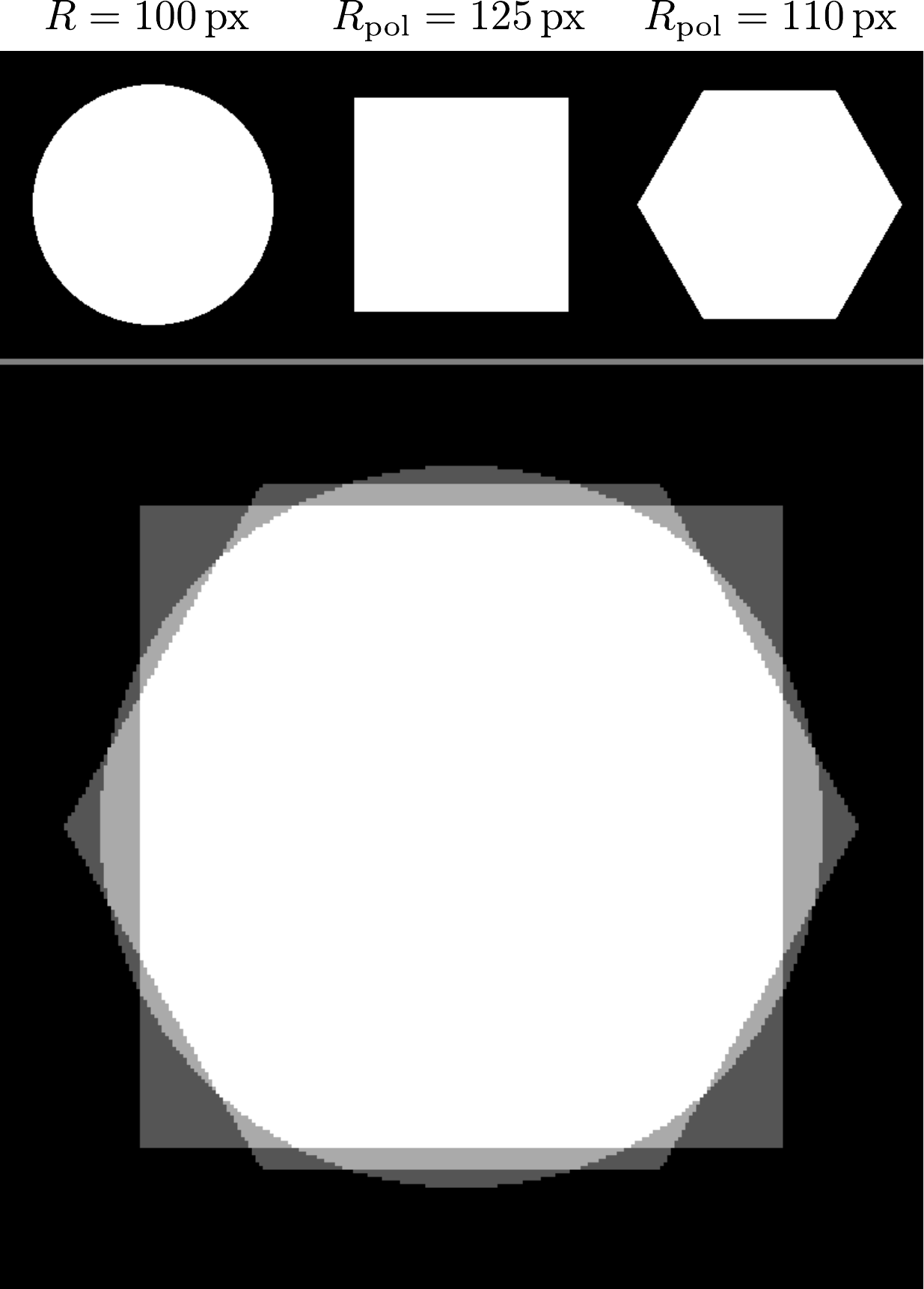}
    \caption{ Example of three figures constructed from a given radius ($R=100~\text{px}$ in this example). Upper panels: each figure shown separately. For the polygonal shapes, their radii $R_{\text{pol}}$ are computed using Eq.~(\ref{eq:R_pol}). Lower panel: comparison of the three figures by superimposition. Up to rounding errors and pixelation, every shape encloses an area of $10000\pi~\text{px}^2$.} 
    \label{fig:Fig2}
\end{figure}

It is important to note that a regular polygon with radius $R$ will always have a smaller area than that of a circle of the same radius (considering the radius of a regular polygon as the distance between its center and any of its vertices). As the  purpose of this work is to compare the same method using different types of illuminated regions, it becomes necessary to build figures with the same illuminated area over the sample in order to avoid a bias towards circles. For this reason, radii for polygons ($R_{\text{pol}}$) were computed in such a way their areas are the closest possible to the area of a circle of radius $R$. Thus, by imposing the area of the polygon to be equal to $\pi R^2$, we obtain that 
\begin{align}
	R_{\text{pol}} = \sqrt{\frac{2\pi/K}{\sin(2\pi/K)} } R,\label{eq:R_pol}
\end{align}
where $K$ is the number of sides the polygon has ($4$ and $6$ for squares and hexagons, respectively). { In general, $K$ is lower-bounded by 3 (triangles) and has no upper bounds since a polygon may have any number of sides . Moreover, according $K$ moves toward infinity, $R_{pol}$ becomes closer to $R$. Figure~\ref{fig:Fig2} shows and example of a circle of radius equal to $100~\text{px}$ together with a square and a hexagon whose radii $R_{\text{pol}}$ were computed using Eq.~(\ref{eq:R_pol}), ensuring that each shape encloses the same area. }

In (b), instead, we used unconnected regions. There were modelled as a plate with the same size of the sample, containing a number of small transparent squares whose sides have length equal to $\ell$ pixels. These squares are randomly distributed on the object, resembling a QR code (see Figure~\ref{fig:Fig1}, right, for an example). This resembles the array of $N \times N$ pinholes in Ref.~\cite{sidorenko2016}, but in this study there is more than one plate and the positions of orifices is random. In order to compare the results with the ones attainable from the aforementioned convex figures of radius $R$, the number of squares is also adjusted in such a way the total transparent area in each $A_j(\vec{r})$ is the closest possible to the one of a circle of radius $R$. Consequently, every illumination function $A_j(\vec{r})$ has $\lfloor\pi R^2/\ell^2\rceil$ squares.  
	
\begin{figure}[!tb]
	\centering
	\includegraphics[width=\columnwidth]{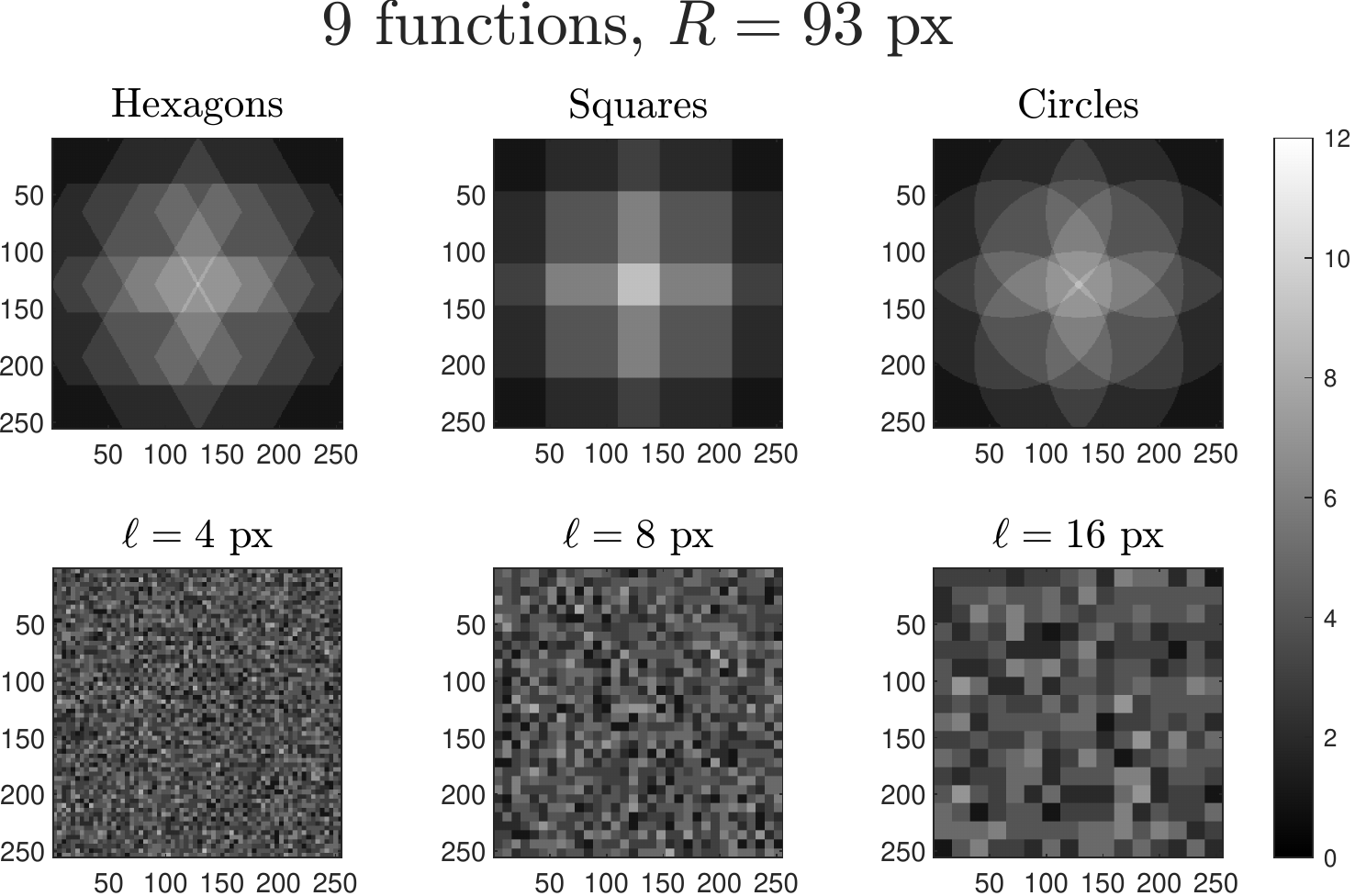}\\[5mm]
	\includegraphics[width=\columnwidth]{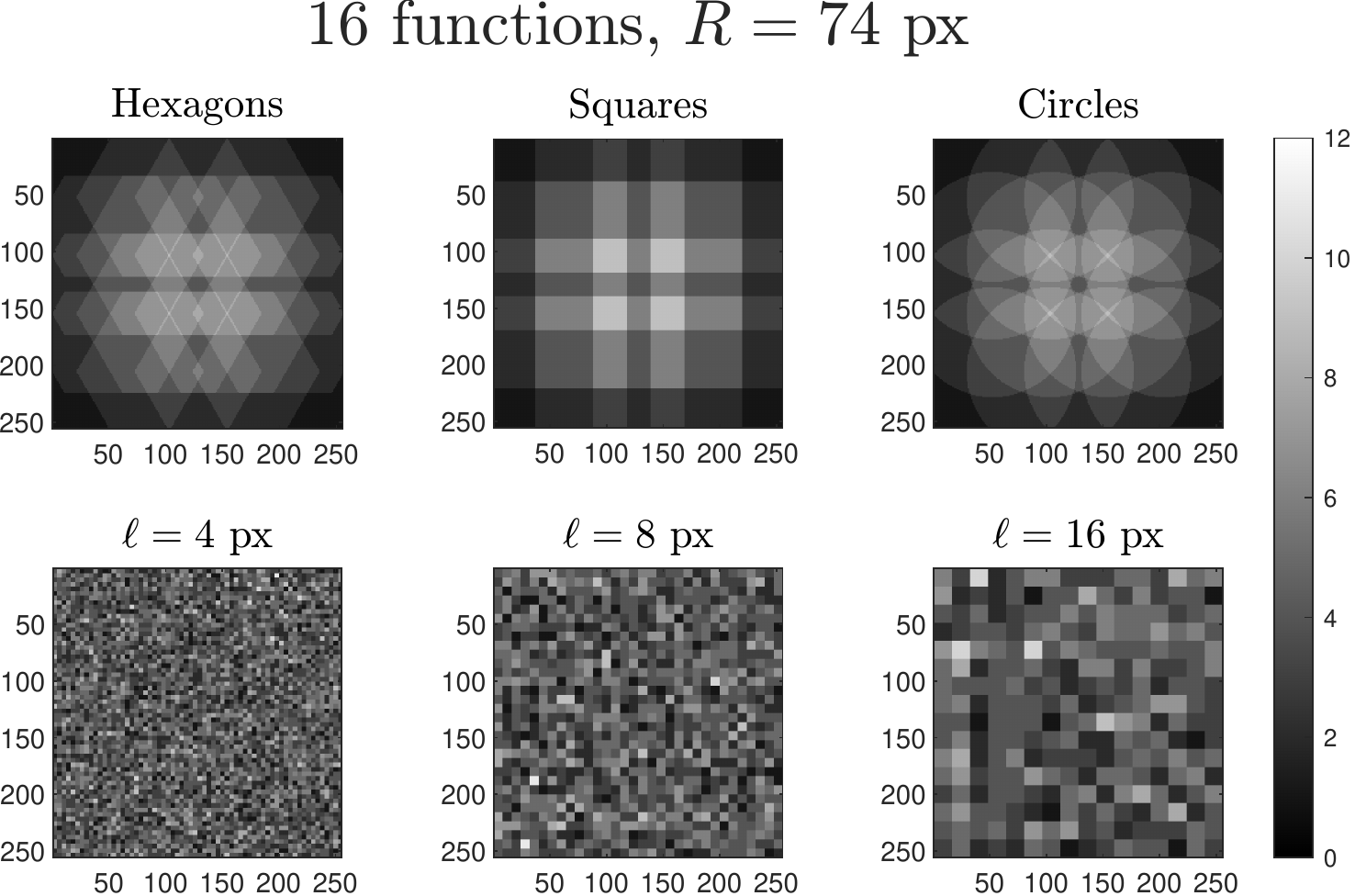}\\[5mm]
	\includegraphics[width=\columnwidth]{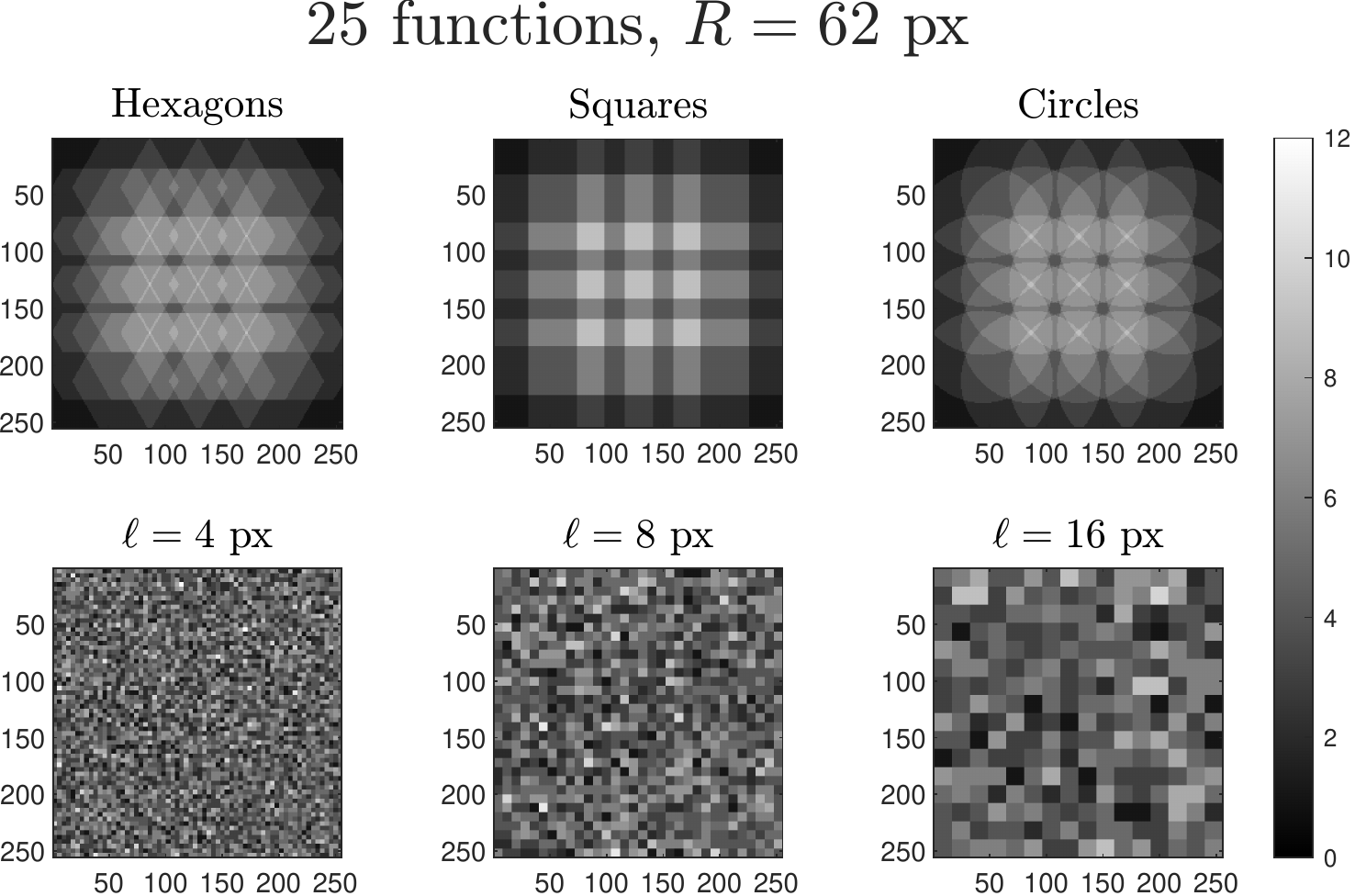}
	\caption{ Superposition of illumination functions for the three cases under study: $N=9$, $16$, and $25$ illumination functions. For each case, the radius was adjusted to the minimum one that allows the complete image to be illuminated at least once. \label{fig:Fig3}}
\end{figure}

{ Figure~\ref{fig:Fig3} shows a superposition of all illumination functions for every shape used in this work. For this work, we considered $N=9$, $16$, and~$25$ illumination functions $\mathcal{S}(\{A_j\})$. Each value of $R$ was adapted in order to be the smallest possible subject to have the complete image illuminated at least once. Consequently, we tested 9 functions with $R=93~\text{px}$, 16 functions with $R=74~\text{px}$, and 25 functions with $R=62~\text{px}$. These minimal radii were chosen in order to make the overlaps between different values of $N$ more uniform. }
 
\par
\begin{figure}[!tbh]
	\centering
	\includegraphics[width=\columnwidth]{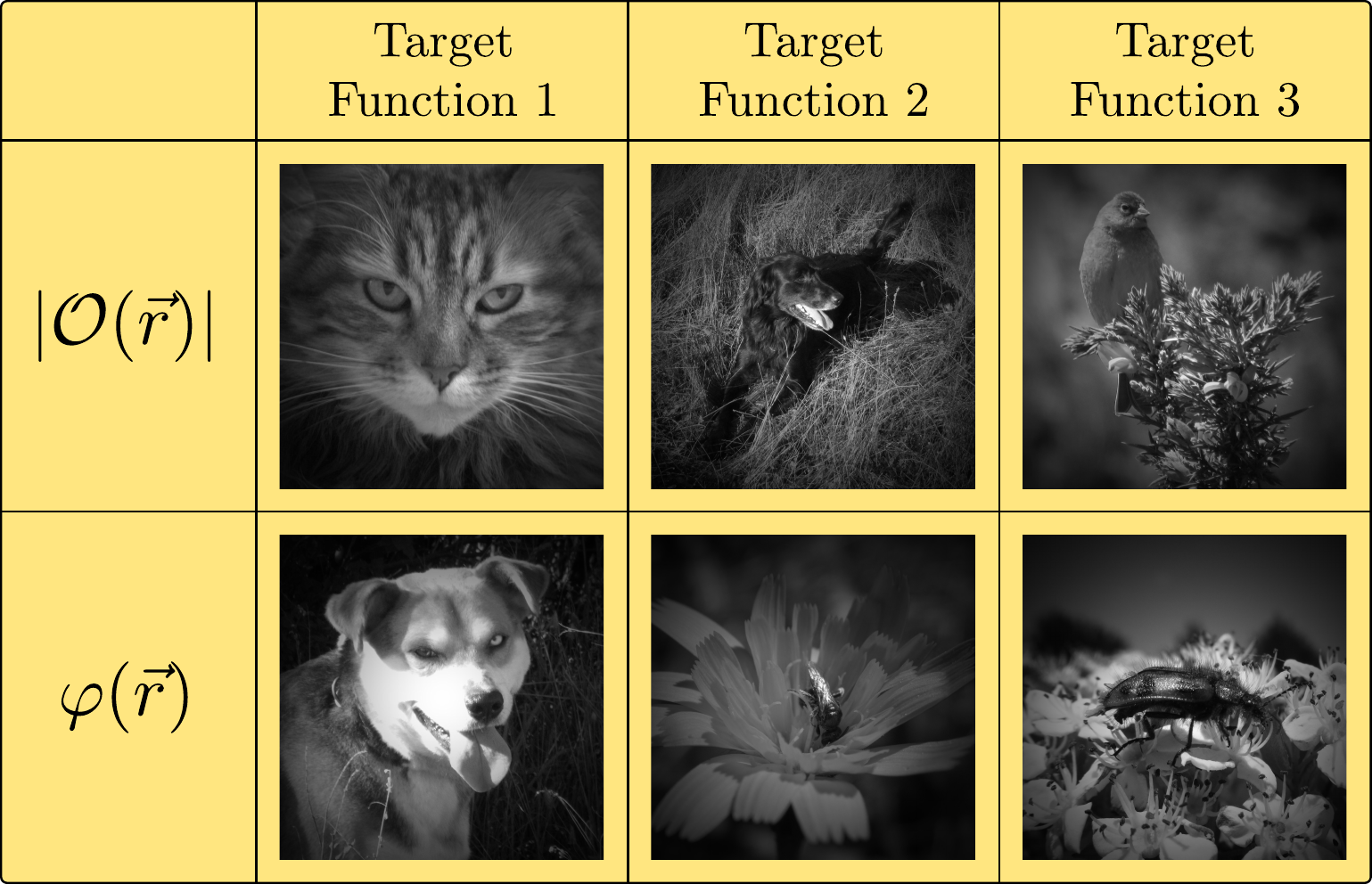}
	\caption{Amplitude and phase of the functions $O(\vec{r})$ used as target functions. The size of each image, in pixels, is $256\times 256$.}
	\label{fig:Fig4}
\end{figure}

 In this study, optical objects are described by transmission functions $\mathcal{O}(\vec{r})$, which are also known as target functions since the reconstruction algorithm must aim to reconstruct a function like those. The target function is built from two images of $ 256 \times 256  $ pixels each, one of them will be used for the amplitude and the other one for the phase, so the object will be described by a transmission function given by
\begin{align}
	\mathcal{O}(\vec{r}) = |\mathcal{O}(\vec{r})|\,e^{2\pi i\varphi(\vec{r})}.
\end{align}
To prevent dependency on the use of the same images, we have selected three different target functions, constructed from different images in grayscale, as Figure~\ref{fig:Fig4} shows. These grayscale values are used to encode values between 0 (black) and 1 (white). The images being used were chosen because they have diverse features that are useful for testing the algorithms: thick and thin stripes; coarse and fine details; well-focused and blurred backgrounds; high and low contrast. 

\subsection{Reconstructed functions and initial guessed function \label{sec:reconst}}

\par In ptychography, one may set a fixed number of iterations for the algorithm to run on, or to define stopping criteria. As our goal is to compare performance between several choices of parameters, we decided to use the same number of iterations for every of the possible shapes being used as illumination functions regardless of the target function. Thus, 200 iterations were used. 

\par Let $n$ be the number of iterations the algorithm has reached, with $1\leq n\leq 200$. After $n$ iterations of the PIE algorithm, a reconstructed function $O_{\text{g},n}\left( \vec{r} \right)$ will be obtained. For this purpose, PIE starts with an initial guessed function $O_{\text{g},0} \left( \vec{r} \right)$ which can be defined, for instance, as a random function or as a constant function. Naturally, it may happen that $O_{\text{g},n}\left( \vec{r} \right)$ exhibits an implicit dependence on the choice of the initial guessed function $O_{\text{g},0} \left( \vec{r} \right)$ and, consequently, the quality of the reconstruction may be strongly conditioned by such a choice. For this reason, we ran the PIE 50 times for every target function and for every shape of illumination functions, each time using a different initial guessed function. 

\subsection{Overview of the algorithm and figures of merit \label{sec:FOM}}

\par Let us recall Figure~\ref{fig:Fig1}. For every illumination function $A_j(\vec{r})$, the transmitted electromagnetic field will be described by $\mathcal{O}(\vec{r})A_j(\vec{r})$. In an experimental situation, a detection system will be able to retrieve intensity distributions $I_j(\vec{u})$ from the Fourier plane, where $\vec{u}$ is the transverse position vector in the Fourier plane. These distributions are the experimental inputs the algorithm needs. In our case, these $I_j(\vec{u})$ are computed via FFT. Once the initial guessed function is defined, the algorithm may start.

\begin{algorithm*}[!htb]
\caption{Summary of PIE algorithm, as shown in Refs.~\cite{rodenburg2004,faulkner2005,rodenburg2007} and in the way it was used on this work. $\mathcal{N}$ is the number of PIE iterations, which is equal to $200$ along this report.  \label{alg:ptycho}}
\begin{algorithmic}[1]
\Procedure{PIE}{$\{I_j(\vec{u})\}_{j=1}^{N}~,~\{A_j(\vec{r})\}_{j=1}^{N}$}       
    \State Define $\displaystyle O_{\text{g},0}^{~}(\vec{r})$	\Comment{This is the initial guessed function}
	\For{$n=1,\dots,\mathcal{N}$}		\Comment{Loop along PIE iterations}
	 	\State $\displaystyle O_{\text{g},n}^{~}(\vec{r}) \gets O_{\text{g},n-1}^{~}(\vec{r})$
		\For{$j=1,\dots,N$}		\Comment{Loop along illumination functions}
			\State $\psi_{\text{g}}(\vec{r}) \gets A_j(\vec{r})\,\displaystyle O_{\text{g},n}^{~}(\vec{r})$
			\State $\Psi_{\text{g}}(\vec{u}) \gets \displaystyle\mathcal{F}\left[\psi_{\text{g}}(\vec{r}) \right](\vec{u})$
			\State $\Psi_{\text{c}}(\vec{u}) \gets \displaystyle\sqrt{I_j(\vec{u})}\, \exp \left( i\,\arg	\Psi_{\text{g}}(\vec{u}) \right)$ 		\Comment{Correct amplitude by using data}
			\State $\psi_{\text{c}}(\vec{r}) \gets \displaystyle\mathcal{F}^{-1}\left[\Psi_{\text{c}}(\vec{u}) \right](\vec{r})$ 
			\State $U_j(\vec{r}) \gets \displaystyle\frac{|A_j(\vec{r})|}{\max_{\vec{r}}\left( |A_j(\vec{r})| \right)} \displaystyle\frac{A_j^{\ast}(\vec{r})}{|A_j(\vec{r})|^2 + \delta}$	\Comment{$\delta\sim 10^{-7}$} \label{eq:Uj}
			\State $O_{\text{g},n}^{~}(\vec{r}) \gets \displaystyle O_{\text{g},n}^{~}(\vec{r}) + U_j(\vec{r}) \left( \psi_{\text{c}}(\vec{r}) - \psi_{\text{g}}(\vec{r}) \right)$  \Comment{Update reconstruction}
		\EndFor
		\State Compute figures of merit regarding $O_{\text{g},n}^{~}(\vec{r})$
		\If{(any stopping criterion is met)}
			\State \textbf{return}
		\EndIf
	\EndFor  
\EndProcedure
\end{algorithmic}
\end{algorithm*}

\par Let us also recall that $O_{\text{g},n}(\vec{r})$ is the reconstructed function after $n$ iterations. Our implementation of PIE is mostly based on Refs.~\cite{rodenburg2004,faulkner2005,rodenburg2007} and summarized in Algorithm~\ref{alg:ptycho}. The inputs the algorithm needs are the intensities $I_j(\vec{u})$ from the Fourier plane and the list of illumination functions. 

Before elaborating details about the figures of merit used, it is necessary to define a matrix norm. Particularly, the following definition will be used, 
\begin{align}
	\|B\| = \sqrt{\sum_{p,q\in\mathcal{S}} |B_{p,q}|^2 },
\end{align}
{ where the sum is performed on the pixels comprising the image ($\mathcal{S}$) to be reconstructed. }

\par Two parameters were used to assess performance: convergence and accuracy. Convergence~($\Delta$) is studied in terms of the difference between the last two estimated functions for each iteration. This parameter should decrease with increasing iterations, as after each iteration these functions should become similar. This parameter is given by 
\begin{align}\label{eq:convergence}
	\Delta =  \frac{ \left\| \kappa O_{\text{g},n}^{~} \left( \vec{r} \right) - O_{\text{g},n-1}^{~} \left( \vec{r} \right) \right\|  } { \sqrt{ \left\| \kappa O_{\text{g},n}^{~} \left( \vec{r} \right) \right\| \left\| O_{\text{g},n-1}^{~} \left( \vec{r}\right)   \right\|}},
\end{align}
where the denominator has been included as normalization factor in order to avoid image size dependence { and to address $\Delta$ as a relative-difference coefficient}. As the matrix difference in the numerator might be artificially increased by a global phase or a global scaling factor, a proportionality constant $\kappa$ has been included in order to minimize this effect. After an optimization, it is possible to show that 
\begin{align}
	\kappa = \frac{\displaystyle\sum_{p,q\in\mathcal{S}} \left(O_{\text{g},n}^{~}\right)_{p,q}^\ast \left(O_{\text{g},n-1}^{~}\right)_{p,q}}{\displaystyle\sum_{p,q\in\mathcal{S}} \left(O_{\text{g},n}^{~}\right)_{p,q}^\ast \left(O_{\text{g},n}^{~}\right)_{p,q}}, \label{eq:kappa}
\end{align}
is the value that assures a minimum of the numerator of Eq.~(\ref{eq:convergence}) with respect to global scaling factors. On the other hand, accuracy~($d$) is studied in terms of the difference between the last estimated function and the target function. This parameter indicates the quality of the retrieval, as it indicates how much the $n$th estimated function resembles the target function: 
\begin{align}\label{eq:accuracy}
	d =  \frac{ \left\| \mathcal{O}\left( \vec{r} \right) - \mu O_{\text{g},n}^{~} \left( \vec{r} \right) \right\|  } { \sqrt{ \left\| \mathcal{O} \left( \vec{r} \right) \right\| \left\| \mu O_{\text{g},n}^{~} \left( \vec{r}\right)   \right\|}},~~~~~~~~
	\mu = \frac{\displaystyle\sum_{p,q\in\mathcal{S}} \left(O_{\text{g},n}^{~}\right)_{p,q}^\ast \left(\mathcal{O}\right)_{p,q}}{\displaystyle\sum_{p,q\in\mathcal{S}} \left(O_{\text{g},n}^{~}\right)_{p,q}^\ast \left(O_{\text{g},n}^{~}\right)_{p,q}}, 
\end{align}
where a normalization factor has also been included here to avoid image size dependence { and to address $d$ also as a relative-difference coefficient}. A proportionality constant $\mu$ was included to remove effects of global phases or global scaling factors as well. This scaling factor $\mu$ was computed in an analogous way as with $\kappa$ in Equation~(\ref{eq:kappa}). For both $d$ and $\Delta$, the closer to zero they are, the better the performance is. If a little abuse of terminology is tolerated, we may name $\Delta$ directly as convergence, and $d$ as accuracy throughout this document. { The use of multiplicative constants, such as $\kappa$ and $\mu$, to avoid the effect of global phases was already proposed in~\cite{fienup1997}. Both $\Delta$ and $d$ are used in this work to assess the performance of the method and the use of every geometry. In an experimental situation, convergence can be used also as a stopping criterion. Accuracy, on the other hand, is not usable in most experimental situations, but rather a figure of merit that can be used mostly for assessing algorithmic performance. }

\subsection{Finite-sized pixels and noise}

\begin{figure*}
    \centering
    \includegraphics[width=0.9\textwidth]{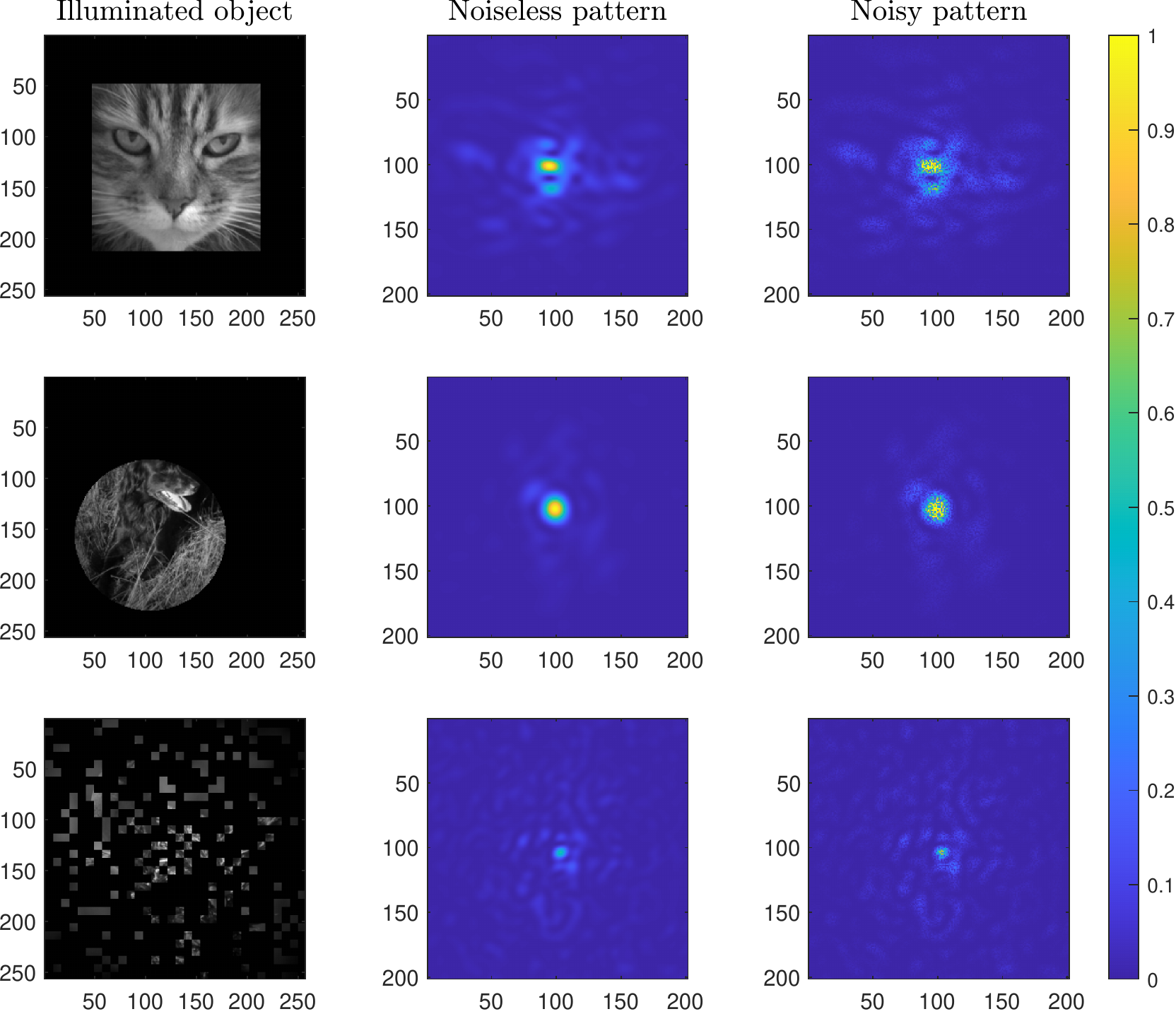}
    \caption{ Example of illumination functions acting on different target functions, their noiseless diffraction pattern and the noisy diffraction pattern. Upper row: $N=9$ and $R=93~\text{px}$ using square illumination functions. Middle row: $N=16$ and $R=74~\text{px}$ using circular illumination functions. Lower row: $N=25$ and $R=62~\text{px}$ using discontinuous illumination functions with $\ell=8~\text{px}$.}
    \label{fig:Fig5}
\end{figure*}

{ Finally, we set up physical parameters in order to include effects from a realistic experimental scenario. Firstly, we now consider the finite size of the detector that can be used in an experiment. That is, the fact that a CCD/CMOS pixel is not exactly a pointlike detector, but rather a small bucket detector capturing light over the complete area each pixel covers. For this reason, although the objects we aim to reconstruct are $256\times 256$, we increased the number of points each FFT/IFFT uses in order to integrate over each pixel. That is, for each illumination function $A_j(\vec{r})$, we have an expected field intensity $\mathcal{I}_j(\vec{r})$ and a expected retrieved distribution $I_j(\vec{r})$. We used 16 points to model each CCD/CMOS pixel. The expected pictures were computed by integrating the expected field intensity over each pixel. For this experimental-case simulation, we considered camera pixels $3.45~\mu\text{m}$-wide and the object to be composed by $8.00~\mu\text{m}$-wide pixels, so the object is, approximately, 2.05 mm-wide. The Fourier transform is performed by a lens with focal length $f=0.1~\text{m}$ in our simulation and we considered illumination from a coherent monochromatic light source of 565.25~nm as wavelength. Thus, the object (modeled as a $256\times256$-sized matrix) was padded with zeros in order to obtain a $8192\times 8192$ matrix. Thus, each expected field intensity ($\mathcal{I}_j$) was a $8192\times 8192$ matrix and each expected retrieved distribution ($I_j$) was a $2048\times 2048$ matrix.  

\par Secondly, we incorporated noise to each expected retrieved distribution. For this purpose, for each shape under consideration, we normalized each dataset $\{I_j\}_{j=1}^{N}$ such that their maximum is equal to 1. Afterwards, speckle noise was added to each $I_j$. We tested speckle noise variances equal to 0 (noiseless case) and 0.20. 

\par Figure~\ref{fig:Fig5} shows some samples of the object (target functions of Figure~\ref{fig:Fig4}) under partial illumination, the expected ideal intensity distribution (noiseless) and one obtained after having applied speckle noise with variance equal to 0.20. Since the diffraction patterns are normalized to have a maximum value of 1, a value of 0.20 as noise variance seems to be relevant.  

}

\section{Results}
\label{sec:results}

For a better comparison of the results both parameters have been plotted over the number of iterations performed in the algorithm, stopping the algorithm after $\mathcal{N}=200$ iterations. Additionally, as aforementioned, each case was studied with 50 choices of initial guessed functions. Thus, our results show bands comprising the central $95\%$ of the results surrounding the mean values of the 50 first guesses. This selection has been made in order to avoid the effect of outliers in our conclusions. As the computational demand increased largely due to the size of the matrices under consideration, we resorted to Single-precision floating-point arithmetic for the computations. As such, we would expect the convergence to end, at best, around $10^{-7}$ since the machine epsilon for single-precision floating-point format is approximately $1.1921\times 10^{-7}$ for Matlab/Octave.

\begin{figure*}[!tb]
	\centering
	\includegraphics[width=\textwidth]{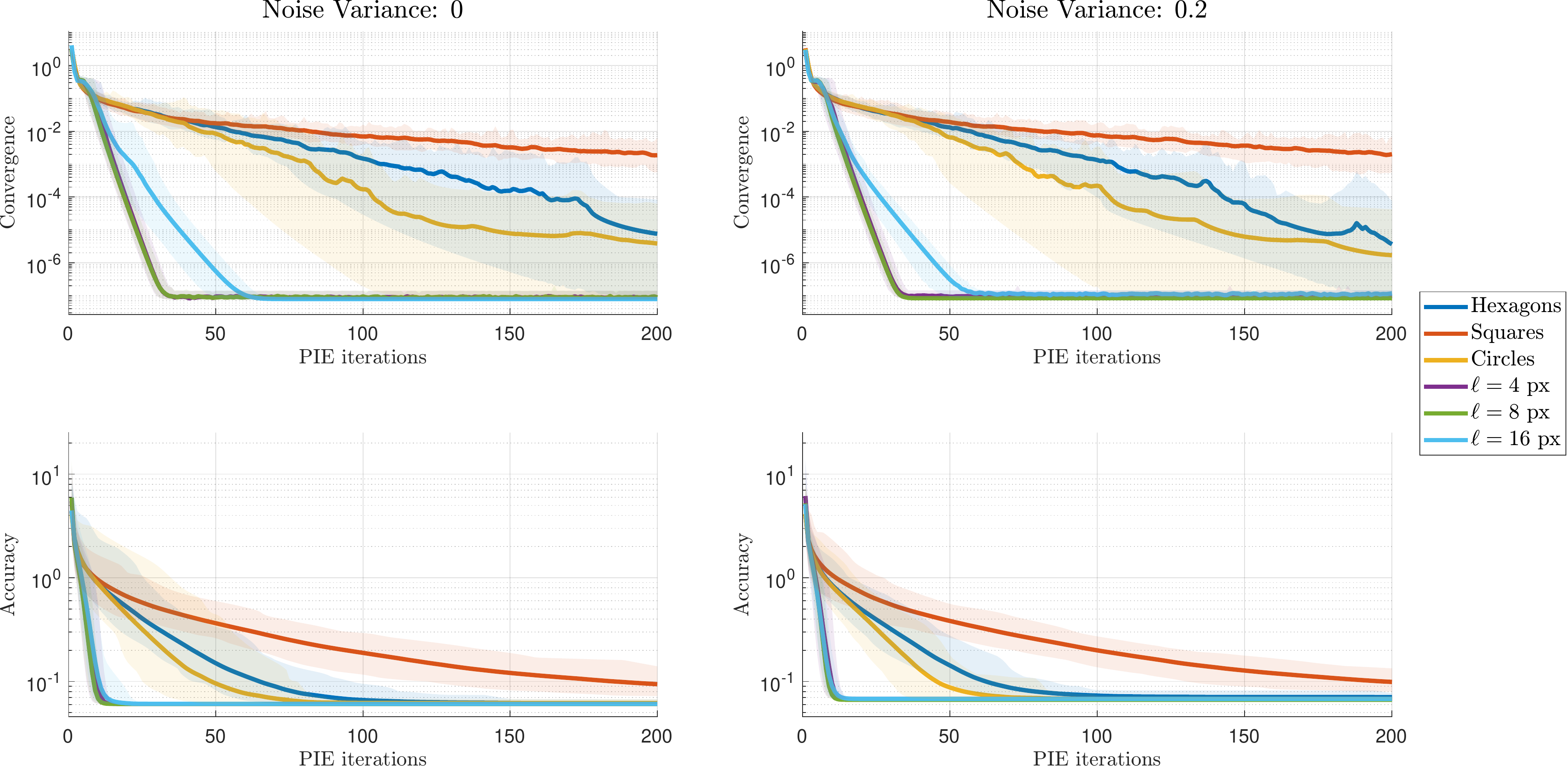}
	\caption{ Convergence (upper panels) and accuracy (lower  panels) obtained for reconstruction of target function~1 using $N=9$ illumination functions and $R=93~\text{px}$. Thick lines represent average over the 50 times the algorithm was applied on different initial guesses. The bands represent the central $95\%$ of the results. The method, regardless of the shape employed, seems to be noise-resistant. However, the use of discontinuous shapes seems to outperform the use of continuous shapes in terms of convergence and necessary PIE iterations.\label{fig:Fig6}}
\end{figure*}

For starters, Figure~\ref{fig:Fig6} shows the results achieved when the algorithm reconstructed target function~1, using $N=9$ illumination functions and $R=93~\text{px}$. It can be seen that the convergence attained by convex figures is dwarfed by the one attained by unconnected regions, which converge much faster. The accuracy reaches to final values much faster when discontinuous shapes are used instead of convex ones: less than 20 PIE iterations using discontinuous shapes lead to the same accuracy that continuous shapes achieve after more than 70 PIE iterations. Moreover, for both figures of merit, convex shapes exhibit great dependence in terms of the first guess. Instead, the choice of the first guess seems to be completely irrelevant when discontinuous shapes are used.

\begin{figure*}[!tb]
	\centering
	\includegraphics[width=\textwidth]{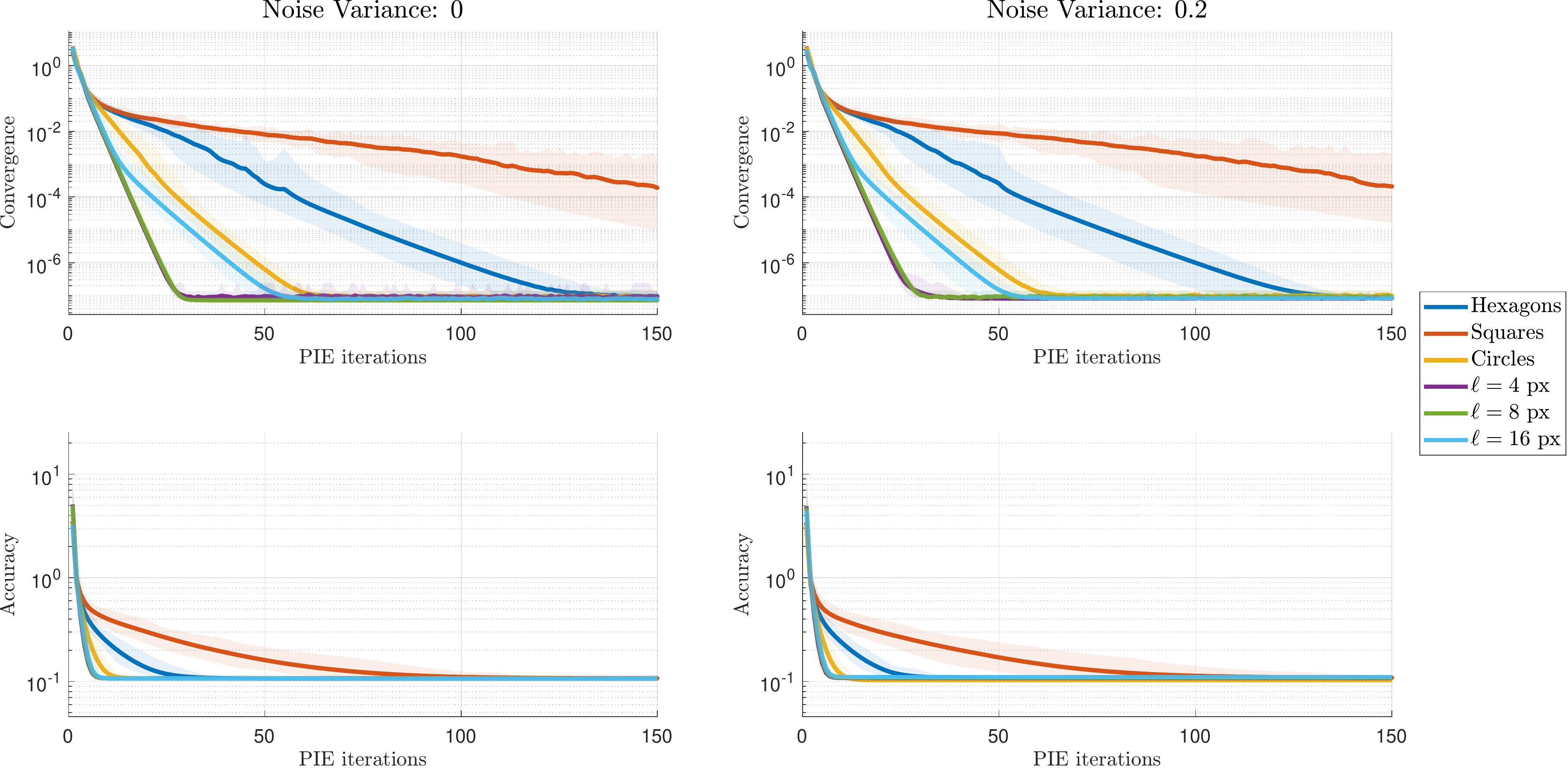}
	\caption{ Convergence (upper panels) and accuracy (lower  panels) obtained for reconstruction of target function~2 using $N=16$ illumination functions and $R=74~\text{px}$. For this configuration, circularly-shaped illumination vastly outperforms polygonally-shaped illumination. Discontinuous illuminations still show better results than convex illuminations, although the advantages are not so evident now. To ease observations, only the first 150 PIE iterations are shown. \label{fig:Fig7}}
\end{figure*}

\par Figure~\ref{fig:Fig7} shows our results for reconstruction of target function~2 using $N=16$ illumination functions and $R=74~\text{px}$. The use of more illumination regions, although smaller ones, leads to faster results (in terms of PIE iterations needed) when compared with the previous case of a smaller number of larger illumination regions. In terms of noise, all shapes seem to be very noise-resistant, but the results from convex regions still depend very strongly on the choice of the first guess---although in a lesser degree than the one observed for $N=9$. Circles now exhibit a performance comparable to the one attained with discontinuously-shaped illumination. These results and the previous ones indicate that, among continuous shapes, circles exhibit the best results. For discontinuous shapes, $\ell=4~\text{px}$ and $\ell=8~\text{px}$ perform almost identically. 

\begin{figure*}[!tb]
	\centering
	\includegraphics[width=\textwidth]{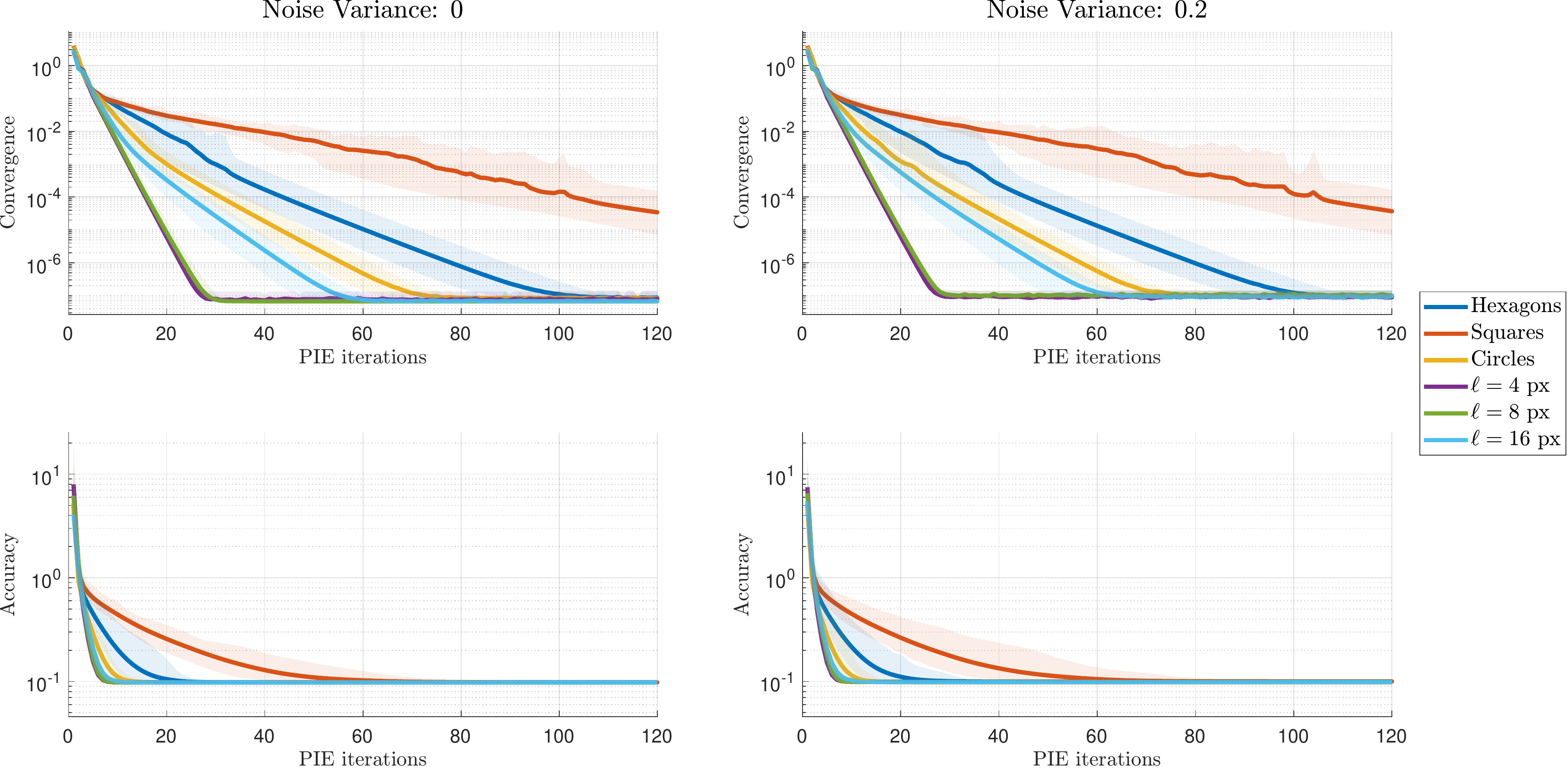}
	\caption{ Convergence (upper panels) and accuracy (lower  panels) obtained for reconstruction of target function~3 using $N=25$ illumination functions and $R=62~\text{px}$. To ease observations, only the first 120 PIE iterations are shown. Regarding convergence and the number of iterations needed, polygonally-shaped illuminations still underperform circularly-shaped illumination which, in turn, underperforms discontinuous illumination. In terms of convergence, however, only squares seem to underperform the other shapes by a relevant margin. \label{fig:Fig8}}
\end{figure*}

Finally, Figure~\ref{fig:Fig8} shows convergence and accuracy, respectively, when target function~3 is reconstructed using $N=25$ and $R=62~\text{px}$. Although the illumination is more uniform in this case (see Figure~\ref{fig:Fig3}), convergence is now slower for circles: they needed around 60 PIE iterations to reach a final result (convergence) when $N=16$, but need almost 80 PIE iterations when $N=25$. Discontinuous shapes with $\ell=16~\text{px}$ also decreased their performance when noise is present: from less than 60 PIE iterations in $N=16$ to almost 70 in $N=25$. On the other hand, $\ell=4~\text{px}$ and $\ell=8~\text{px}$ perform almost identically in every configuration, needing around 30 PIE iterations regardless the value of $N$.

Accuracy, unlike convergence, seems to benefit from increasing the number of illumination functions regardless of the shrinking radii. On one hand, this would be expected since ptychography may be used for achieving superresolution as consequence of information redundancy~\cite{rodenburg2007,maiden2011}. It is natural to think more redundancy would lead to better performance and these  results seem to agree with that. On the other hand, convergence might need a trade-off between the number of illumination functions used and their width. This is not completely unexpected since Ref.~\cite{bunk2008} already showed that performance is non-monotonically linked to the overlap between illumination functions. Although it is easy to quantify an overlap between two functions, a study on the overlap between $N$ functions for different radius and shapes lies beyond the scope of the current work and can be addressed in a future study.

\begin{figure*}
  \begin{subfigure}{0.31\textwidth}
    \includegraphics[width=\linewidth]{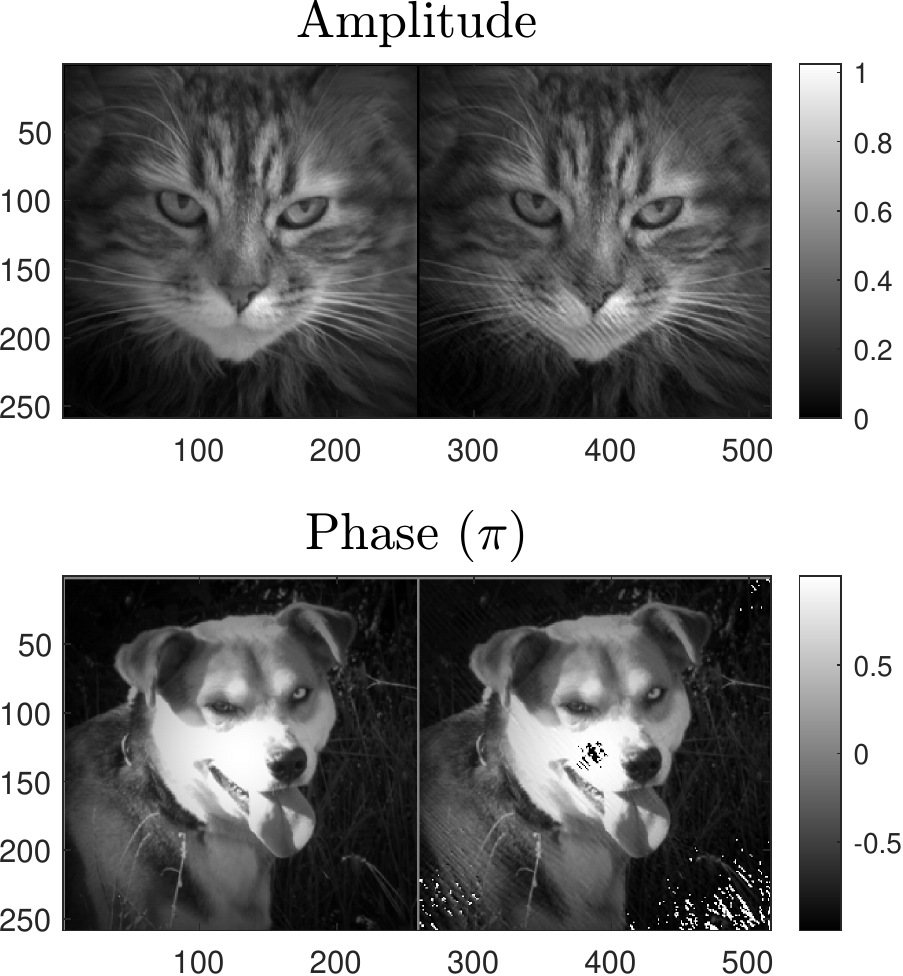}
    \caption{Target function 1, using squares (${N=9}, {R=93~\text{px}}$).} \label{fig:Fig9a}
  \end{subfigure}%
  \hspace*{\fill}   
  \begin{subfigure}{0.31\textwidth}
    \includegraphics[width=\linewidth]{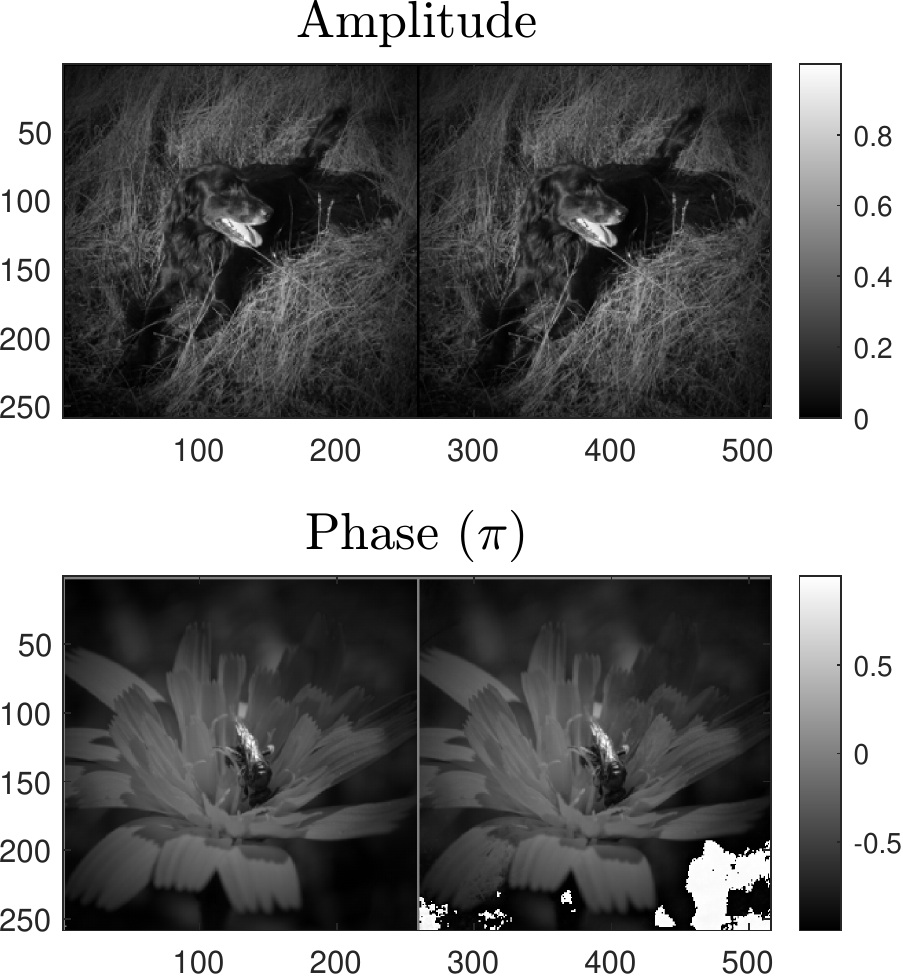}
    \caption{Target function 2, using circles (${N=16}, {R=74~\text{px}}$).} \label{fig:Fig9b}
  \end{subfigure}%
  \hspace*{\fill}   
  \begin{subfigure}{0.31\textwidth}
    \includegraphics[width=\linewidth]{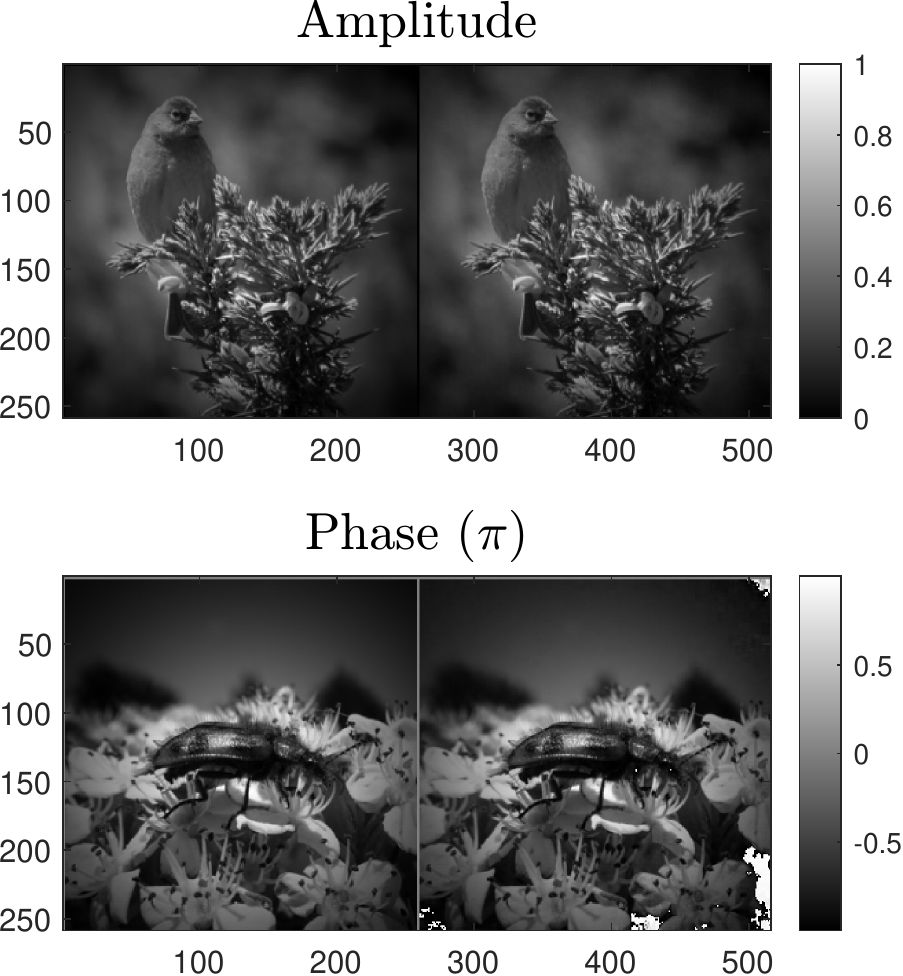}
    \caption{Target function 3, using $\ell=8~\text{px}$ (${N=25}, {R=62~\text{px}}$).} \label{fig:Fig9c}
  \end{subfigure}
\caption{ Some of the reconstructed figures using different illumination shapes. For every pair of images, left one represents the target function. Right one is the result of ptychographic reconstruction. They correspond to the same configurations exemplified in Figure~\ref{fig:Fig5}. } \label{fig:Fig9}
\end{figure*}

All results indicate that unconnected regions constructed from smaller squares perform better than the ones built from larger squares. A possible explanation lies in the fact that, for a fixed area, the smaller squares gather information from a more diverse set of regions on the image than convex shapes. In order to appreciate the results of ptychographic reconstruction, Figure~\ref{fig:Fig9} shows some reconstructed images compared with their respective target functions. Figure~\ref{fig:Fig9a} shows the result of a reconstruction using large squares ($N=9$) on target function 1. One may see a kind of artefact on the reconstructed images which is not seen on the other shapes. Remarkably, Figure~\ref{fig:Fig7} showed squares struggled to converge. Figure~\ref{fig:Fig9b} shows reconstruction after using circles ($N=16$) on target function 2, leading to good results. Finally, Figure~\ref{fig:Fig9c} shows reconstruction of target function 3 after using discontinuous illumination functions ($N=25, \ell=8~\text{px}$). leading to seemingly high-quality results.

\section{Conclusion }
\label{sec:conclusion}

We used three different complex target functions that exhibit several diverse features (coarse and fine details, high and low contrast, etc) in order to test the scope of our conclusions. Additionally, as it could be expected, the use of more illumination functions leads to more experimental information, which leads to better results in terms of the number of iterations needed to achieve a final accuracy level. { However, and perhaps unexpectedly, the use of unconnected illumination functions resembling QR codes seems to outperform the use of regular shapes by a substantial margin in some cases. Moreover, the effect of speckle noise for these shapes seemed not so relevant. In this regard, ptychography might be very noise-resistant against speckle noise regardless of the shape of the illumination functions.  }

The results suggest that gathering information from more diverse regions on the image lead to better results. For this reason, unconnected regions built from smaller squares performed better than the ones produced from larger squares, which, in turn, outperformed convex shapes. The use of unconnected regions is something it could be implemented through the current technology of spatial light modulators if visible light is used. However, we acknowledge smaller squares might be experimentally more challenging to implement than larger ones. As the advantages were more relevant for smaller number of illuminations used, these shapes could be used when data acquisition must be done very quickly, like in a biologically active sample.  

We anticipate these results can be useful for any topic in which ptychography has been used, mostly microscopy, as well as in any field in which high-resolution imaging is not only necessary, but difficult, such as observational astronomy.

\begin{backmatter}

\bmsection{Funding}
DIUFRO grant DI20-0154

\bmsection{Acknowledgments}
F.S. acknowledges partial financial support from the Master of Science in Physics program at Universidad de La Frontera. M.A.S.-P. acknowledges funding from Universidad de La Frontera through DIUFRO grant DI20-0154. The authors also would like to thank Fabi\'an Torres Ruiz and Leonardo Teixeira Neves for fruitful conversations. 

\bmsection{Disclosures}
The authors declare no conflicts of interest.

\bmsection{Data availability} 
Data underlying the results presented in this paper are not publicly available at this time but may be obtained from the authors upon reasonable request.

\end{backmatter}

\bibliography{refs,ref2}

\begin{thebibliography}{10}
\newcommand{\enquote}[1]{``#1''}

\bibitem{zernike1942}
F.~Zernike, \enquote{Phase contrast, a new method for the microscopic
  observation of transparent objects,} {\protect\JournalTitle{Physica}}
  \textbf{9}, 686--698 (1942).

\bibitem{shaked2013}
N.~T. Shaked, Z.~Zalevsky, and L.~L. Satterwhite, eds., \emph{Biomedical
  Optical Phase Microscopy and Nanoscopy} ({Academic Press}, {Oxford}, 2013),
  1st ed.

\bibitem{rodenburg1989}
J.~Rodenburg, \enquote{The phase problem, microdiffraction and
  wavelength-limited resolution \textemdash{} a discussion,}
  {\protect\JournalTitle{Ultramicroscopy}} \textbf{27}, 413--422 (1989).

\bibitem{rempfer1992}
G.~F. Rempfer and O.~Hayes~Griffith, \enquote{Emission microscopy and related
  techniques: {{Resolution}} in photoelectron microscopy, low energy electron
  microscopy and mirror electron microscopy,}
  {\protect\JournalTitle{Ultramicroscopy}} \textbf{47}, 35--54 (1992).

\bibitem{coene1992}
W.~Coene, G.~Janssen, M.~{Op de Beeck}, and D.~Van~Dyck, \enquote{Phase
  retrieval through focus variation for ultra-resolution in field-emission
  transmission electron microscopy,} {\protect\JournalTitle{Physical Review
  Letters}} \textbf{69}, 3743--3746 (1992).

\bibitem{opdebeeck1996}
M.~{Op de Beeck}, D.~Van~Dyck, and W.~Coene, \enquote{Wave function
  reconstruction in {{HRTEM}}: The parabola method,}
  {\protect\JournalTitle{Ultramicroscopy}} \textbf{64}, 167--183 (1996).

\bibitem{allen2000}
L.~J. Allen, H.~M.~L. Faulkner, and H.~Leeb, \enquote{Inversion of dynamical
  electron diffraction data including absorption,} {\protect\JournalTitle{Acta
  Crystallographica Section A: Foundations of Crystallography}} \textbf{56},
  119--126 (2000).

\bibitem{allen2001a}
L.~J. Allen, H.~M.~L. Faulkner, M.~P. Oxley, and D.~Paganin, \enquote{Phase
  retrieval and aberration correction in the presence of vortices in
  high-resolution transmission electron microscopy,}
  {\protect\JournalTitle{Ultramicroscopy}} \textbf{88}, 85--97 (2001).

\bibitem{millane1990}
R.~P. Millane, \enquote{Phase retrieval in crystallography and optics,}
  {\protect\JournalTitle{JOSA A}} \textbf{7}, 394--411 (1990).

\bibitem{fitzgerald2000}
R.~Fitzgerald, \enquote{Phase-{{Sensitive X}}-{{Ray Imaging}},}
  {\protect\JournalTitle{Physics Today}} \textbf{53}, 23--26 (2000).

\bibitem{taylor2003}
G.~Taylor, \enquote{The phase problem,} {\protect\JournalTitle{Acta
  Crystallographica Section D: Biological Crystallography}} \textbf{59},
  1881--1890 (2003).

\bibitem{lewis2004}
R.~A. Lewis, \enquote{Medical phase contrast x-ray imaging: Current status and
  future prospects,} {\protect\JournalTitle{Physics in Medicine and Biology}}
  \textbf{49}, 3573--3583 (2004).

\bibitem{wu2005}
X.~Wu, H.~Liu, and A.~Yan, \enquote{X-ray phase-attenuation duality and phase
  retrieval,} {\protect\JournalTitle{Optics Letters}} \textbf{30}, 379--381
  (2005).

\bibitem{burvall2011}
A.~Burvall, U.~Lundstr{\"o}m, P.~A.~C. Takman, D.~H. Larsson, and H.~M. Hertz,
  \enquote{Phase retrieval in {{X-ray}} phase-contrast imaging suitable for
  tomography,} {\protect\JournalTitle{Optics Express}} \textbf{19},
  10359--10376 (2011).

\bibitem{wu2022}
Y.~Wu, L.~Zhang, S.~Guo, L.~Zhang, F.~Gao, M.~Jia, and Z.~Zhou,
  \enquote{Enhanced phase retrieval via deep concatenation networks for in-line
  {{X-ray}} phase contrast imaging,} {\protect\JournalTitle{Physica Medica}}
  \textbf{95}, 41--49 (2022).

\bibitem{gerchberg1972}
R.~W. Gerchberg and W.~O. Saxton, \enquote{A practical algorithm for the
  determination of phase from image and diffraction plane pictures,}
  {\protect\JournalTitle{Optik}} \textbf{35}, 237--246 (1972).

\bibitem{fienup1978}
J.~R. Fienup, \enquote{Reconstruction of an object from the modulus of its
  {{Fourier}} transform,} {\protect\JournalTitle{Optics Letters}} \textbf{3},
  27 (1978).

\bibitem{fienup1982}
J.~R. Fienup, \enquote{Phase retrieval algorithms: A comparison,}
  {\protect\JournalTitle{Applied Optics}} \textbf{21}, 2758 (1982).

\bibitem{bauschke2002}
H.~H. Bauschke, P.~L. Combettes, and D.~R. Luke, \enquote{Phase retrieval,
  error reduction algorithm, and {{Fienup}} variants: A view from convex
  optimization,} {\protect\JournalTitle{JOSA A}} \textbf{19}, 1334--1345
  (2002).

\bibitem{zhao2020}
T.~Zhao and Y.~Chi, \enquote{Modified {{Gerchberg}}\textendash{{Saxton}}
  ({{G-S}}) {{Algorithm}} and {{Its Application}},}
  {\protect\JournalTitle{Entropy}} \textbf{22}, 1354 (2020).

\bibitem{hoppe1982}
W.~Hoppe, \enquote{Trace structure analysis, ptychography, phase tomography,}
  {\protect\JournalTitle{Ultramicroscopy}} \textbf{10}, 187--198 (1982).

\bibitem{faulkner2004}
H.~M.~L. Faulkner and J.~M. Rodenburg, \enquote{Movable {{Aperture Lensless
  Transmission Microscopy}}: {{A Novel Phase Retrieval Algorithm}},}
  {\protect\JournalTitle{Physical Review Letters}} \textbf{93}, 023903 (2004).

\bibitem{rodenburg2004}
J.~M. Rodenburg and H.~M.~L. Faulkner, \enquote{A phase retrieval algorithm for
  shifting illumination,} {\protect\JournalTitle{Applied Physics Letters}}
  \textbf{85}, 4795--4797 (2004).

\bibitem{faulkner2005}
H.~Faulkner and J.~Rodenburg, \enquote{Error tolerance of an iterative phase
  retrieval algorithm for moveable illumination microscopy,}
  {\protect\JournalTitle{Ultramicroscopy}} \textbf{103}, 153--164 (2005).

\bibitem{rodenburg2008}
J.~Rodenburg, \enquote{Ptychography and {{Related Diffractive Imaging
  Methods}},} in \emph{Advances in {{Imaging}} and {{Electron Physics}},}  vol.
  150 ({Elsevier}, 2008), pp. 87--184.

\bibitem{bunk2008}
O.~Bunk, M.~Dierolf, S.~Kynde, I.~Johnson, O.~Marti, and F.~Pfeiffer,
  \enquote{Influence of the overlap parameter on the convergence of the
  ptychographical iterative engine,} {\protect\JournalTitle{Ultramicroscopy}}
  \textbf{108}, 481--487 (2008).

\bibitem{rodenburg2007}
J.~Rodenburg, A.~Hurst, and A.~Cullis, \enquote{Transmission microscopy without
  lenses for objects of unlimited size,}
  {\protect\JournalTitle{Ultramicroscopy}} \textbf{107}, 227--231 (2007).

\bibitem{maiden2011}
A.~M. Maiden, M.~J. Humphry, F.~Zhang, and J.~M. Rodenburg,
  \enquote{Superresolution imaging via ptychography,}
  {\protect\JournalTitle{Journal of the Optical Society of America A}}
  \textbf{28}, 604 (2011).

\bibitem{maiden2009}
A.~M. Maiden and J.~M. Rodenburg, \enquote{An improved ptychographical phase
  retrieval algorithm for diffractive imaging,}
  {\protect\JournalTitle{Ultramicroscopy}} \textbf{109}, 1256--1262 (2009).

\bibitem{maiden2012}
A.~M. Maiden, M.~J. Humphry, and J.~M. Rodenburg, \enquote{Ptychographic
  transmission microscopy in three dimensions using a multi-slice approach,}
  {\protect\JournalTitle{Journal of the Optical Society of America A}}
  \textbf{29}, 1606 (2012).

\bibitem{konijnenberg2016}
A.~Konijnenberg, W.~Coene, S.~Pereira, and H.~Urbach, \enquote{Combining
  ptychographical algorithms with the {{Hybrid Input-Output}} ({{HIO}})
  algorithm,} {\protect\JournalTitle{Ultramicroscopy}} \textbf{171}, 43--54
  (2016).

\bibitem{zheng2013}
G.~Zheng, R.~Horstmeyer, and C.~Yang, \enquote{Wide-field, high-resolution
  {{Fourier}} ptychographic microscopy,} {\protect\JournalTitle{Nature
  Photonics}} \textbf{7}, 739--745 (2013).

\bibitem{yeh2015}
L.-H. Yeh, J.~Dong, J.~Zhong, L.~Tian, M.~Chen, G.~Tang, M.~Soltanolkotabi, and
  L.~Waller, \enquote{Experimental robustness of {{Fourier}} ptychography phase
  retrieval algorithms,} {\protect\JournalTitle{Optics Express}} \textbf{23},
  33214--33240 (2015).

\bibitem{zhanglei-lei2017}
{Zhang Lei-Lei}, {Tang Li-Jin}, {Zhang Mu-Yang}, and {Liang Yan-Mei},
  \enquote{Symmetric illumination in {{Fourier}} ptychography,}
  {\protect\JournalTitle{Acta Physica Sinica}} \textbf{66}, 224201 (2017).

\bibitem{konda2020}
P.~C. Konda, L.~Loetgering, K.~C. Zhou, S.~Xu, A.~R. Harvey, and R.~Horstmeyer,
  \enquote{Fourier ptychography: Current applications and future promises,}
  {\protect\JournalTitle{Optics Express}} \textbf{28}, 9603 (2020).

\bibitem{bianco2021}
V.~Bianco, B.~Mandracchia, J.~Bhal, D.~Barone, P.~Memmolo, and P.~Ferraro,
  \enquote{Miscalibration-{{Tolerant Fourier Ptychography}},}
  {\protect\JournalTitle{IEEE Journal of Selected Topics in Quantum
  Electronics}} \textbf{27}, 1--17 (2021).

\bibitem{lee2021}
H.~Lee, B.~Chon, and H.~Ahn, \enquote{Rapid misalignment correction method in
  reflective fourier ptychographic microscopy for full field of view
  reconstruction,} {\protect\JournalTitle{Optics and Lasers in Engineering}}
  \textbf{138}, 106418 (2021).

\bibitem{nashed2014}
Y.~S.~G. Nashed, D.~J. Vine, T.~Peterka, J.~Deng, R.~Ross, and C.~Jacobsen,
  \enquote{Parallel ptychographic reconstruction,}
  {\protect\JournalTitle{Optics Express}} \textbf{22}, 32082 (2014).

\bibitem{odstrcil2016}
M.~Odstrcil, P.~Baksh, S.~A. Boden, R.~Card, J.~E. Chad, J.~G. Frey, and W.~S.
  Brocklesby, \enquote{Ptychographic coherent diffractive imaging with
  orthogonal probe relaxation,} {\protect\JournalTitle{Optics Express}}
  \textbf{24}, 8360 (2016).

\bibitem{maiden2017}
A.~Maiden, D.~Johnson, and P.~Li, \enquote{Further improvements to the
  ptychographical iterative engine,} {\protect\JournalTitle{Optica}}
  \textbf{4}, 736 (2017).

\bibitem{yao2021}
Y.~Yao, Y.~Jiang, J.~Klug, Y.~Nashed, C.~Roehrig, C.~Preissner, F.~Marin,
  M.~Wojcik, O.~Cossairt, Z.~Cai, S.~Vogt, B.~Lai, and J.~Deng,
  \enquote{Broadband {{X-ray}} ptychography using multi-wavelength algorithm,}
  {\protect\JournalTitle{Journal of Synchrotron Radiation}} \textbf{28},
  309--317 (2021).

\bibitem{thibault2009}
P.~Thibault, M.~Dierolf, O.~Bunk, A.~Menzel, and F.~Pfeiffer, \enquote{Probe
  retrieval in ptychographic coherent diffractive imaging,}
  {\protect\JournalTitle{Ultramicroscopy}} \textbf{109}, 338--343 (2009).

\bibitem{maiden2010}
A.~M. Maiden, J.~M. Rodenburg, and M.~J. Humphry, \enquote{Optical
  ptychography: A practical implementation with useful resolution,}
  {\protect\JournalTitle{Optics Letters}} \textbf{35}, 2585 (2010).

\bibitem{claus2013}
D.~Claus, D.~J. Robinson, D.~G. Chetwynd, Y.~Shuo, W.~T. Pike, J.~J. De~J
  Toriz~Garcia, and J.~M. Rodenburg, \enquote{Dual wavelength optical metrology
  using ptychography,} {\protect\JournalTitle{Journal of Optics}} \textbf{15},
  035702 (2013).

\bibitem{zhang2013}
F.~Zhang, I.~Peterson, J.~{Vila-Comamala}, A.~Diaz, F.~Berenguer, R.~Bean,
  B.~Chen, A.~Menzel, I.~K. Robinson, and J.~M. Rodenburg, \enquote{Translation
  position determination in ptychographic coherent diffraction imaging,}
  {\protect\JournalTitle{Optics Express}} \textbf{21}, 13592 (2013).

\bibitem{godden2014}
T.~M. Godden, R.~Suman, M.~J. Humphry, J.~M. Rodenburg, and A.~M. Maiden,
  \enquote{Ptychographic microscope for three-dimensional imaging,}
  {\protect\JournalTitle{Optics Express}} \textbf{22}, 12513 (2014).

\bibitem{li2019}
{Li}, {Wen}, {Song}, {Jiang}, {Zhang}, {Liu}, and {Wei}, \enquote{Imaging
  {{Correlography Using Ptychography}},} {\protect\JournalTitle{Applied
  Sciences}} \textbf{9}, 4377 (2019).

\bibitem{chang2020}
C.~Chang, X.~Pan, H.~Tao, C.~Liu, S.~P. Veetil, and J.~Zhu,
  \enquote{Single-shot ptychography with highly tilted illuminations,}
  {\protect\JournalTitle{Optics Express}} \textbf{28}, 28441 (2020).

\bibitem{rodenburg2007a}
J.~M. Rodenburg, A.~C. Hurst, A.~G. Cullis, B.~R. Dobson, F.~Pfeiffer, O.~Bunk,
  C.~David, K.~Jefimovs, and I.~Johnson, \enquote{Hard-{{X-Ray Lensless
  Imaging}} of {{Extended Objects}},} {\protect\JournalTitle{Physical Review
  Letters}} \textbf{98}, 034801 (2007).

\bibitem{thibault2008}
P.~Thibault, M.~Dierolf, A.~Menzel, O.~Bunk, C.~David, and F.~Pfeiffer,
  \enquote{High-{{Resolution Scanning X-ray Diffraction Microscopy}},}
  {\protect\JournalTitle{Science}} \textbf{321}, 379--382 (2008).

\bibitem{dierolf2010}
M.~Dierolf, A.~Menzel, P.~Thibault, P.~Schneider, C.~M. Kewish, R.~Wepf,
  O.~Bunk, and F.~Pfeiffer, \enquote{Ptychographic {{X-ray}} computed
  tomography at the nanoscale,} {\protect\JournalTitle{Nature}} \textbf{467},
  436--439 (2010).

\bibitem{dierolf2010a}
M.~Dierolf, P.~Thibault, A.~Menzel, C.~M. Kewish, K.~Jefimovs, I.~Schlichting,
  K.~von K{\"o}nig, O.~Bunk, and F.~Pfeiffer, \enquote{Ptychographic coherent
  diffractive imaging of weakly scattering specimens,}
  {\protect\JournalTitle{New Journal of Physics}} \textbf{12}, 035017 (2010).

\bibitem{edo2013}
T.~B. Edo, D.~J. Batey, A.~M. Maiden, C.~Rau, U.~Wagner, Z.~D. Pe{\v s}i{\'c},
  T.~A. Waigh, and J.~M. Rodenburg, \enquote{Sampling in x-ray ptychography,}
  {\protect\JournalTitle{Physical Review A}} \textbf{87}, 053850 (2013).

\bibitem{maiden2013}
A.~Maiden, G.~Morrison, B.~Kaulich, A.~Gianoncelli, and J.~Rodenburg,
  \enquote{Soft {{X-ray}} spectromicroscopy using ptychography with randomly
  phased illumination,} {\protect\JournalTitle{Nature Communications}}
  \textbf{4}, 1669 (2013).

\bibitem{stockmar2015}
M.~Stockmar, I.~Zanette, M.~Dierolf, B.~Enders, R.~Clare, F.~Pfeiffer,
  P.~Cloetens, A.~Bonnin, and P.~Thibault, \enquote{X-{{Ray Near-Field
  Ptychography}} for {{Optically Thick Specimens}},}
  {\protect\JournalTitle{Physical Review Applied}} \textbf{3}, 014005 (2015).

\bibitem{morrison2018}
G.~R. Morrison, F.~Zhang, A.~Gianoncelli, and I.~K. Robinson, \enquote{X-ray
  ptychography using randomized zone plates,} {\protect\JournalTitle{Optics
  Express}} \textbf{26}, 14915 (2018).

\bibitem{pfeiffer2018}
F.~Pfeiffer, \enquote{X-ray ptychography,} {\protect\JournalTitle{Nature
  Photonics}} \textbf{12}, 9--17 (2018).

\bibitem{holler2019}
M.~Holler, M.~Odstrcil, M.~{Guizar-Sicairos}, M.~Lebugle, E.~M{\"u}ller,
  S.~Finizio, G.~Tinti, C.~David, J.~Zusman, W.~Unglaub, O.~Bunk, J.~Raabe,
  A.~F.~J. Levi, and G.~Aeppli, \enquote{Three-dimensional imaging of
  integrated circuits with macro- to nanoscale zoom,}
  {\protect\JournalTitle{Nature Electronics}} \textbf{2}, 464--470 (2019).

\bibitem{kahnt2021}
M.~Kahnt, L.~Grote, D.~Br{\"u}ckner, M.~Seyrich, F.~Wittwer, D.~Koziej, and
  C.~G. Schroer, \enquote{Multi-slice ptychography enables high-resolution
  measurements in extended chemical reactors,}
  {\protect\JournalTitle{Scientific Reports}} \textbf{11}, 1500 (2021).

\bibitem{nellist1998}
P.~D. Nellist and J.~M. Rodenburg, \enquote{Electron {{Ptychography}}. {{I}}.
  {{Experimental Demonstration Beyond}} the {{Conventional Resolution
  Limits}},} {\protect\JournalTitle{Acta Crystallographica Section A
  Foundations of Crystallography}} \textbf{54}, 49--60 (1998).

\bibitem{haigh2009}
S.~J. Haigh, H.~Sawada, and A.~I. Kirkland, \enquote{Atomic {{Structure Imaging
  Beyond Conventional Resolution Limits}} in the {{Transmission Electron
  Microscope}},} {\protect\JournalTitle{Physical Review Letters}} \textbf{103},
  126101 (2009).

\bibitem{hue2010}
F.~H{\"u}e, J.~M. Rodenburg, A.~M. Maiden, F.~Sweeney, and P.~A. Midgley,
  \enquote{Wave-front phase retrieval in transmission electron microscopy via
  ptychography,} {\protect\JournalTitle{Physical Review B}} \textbf{82}, 121415
  (2010).

\bibitem{hurst2010}
A.~C. Hurst, T.~B. Edo, T.~Walther, F.~Sweeney, and J.~M. Rodenburg,
  \enquote{Probe position recovery for ptychographical imaging,}
  {\protect\JournalTitle{Journal of Physics: Conference Series}} \textbf{241},
  012004 (2010).

\bibitem{humphry2012}
M.~Humphry, B.~Kraus, A.~Hurst, A.~Maiden, and J.~Rodenburg,
  \enquote{Ptychographic electron microscopy using high-angle dark-field
  scattering for sub-nanometre resolution imaging,}
  {\protect\JournalTitle{Nature Communications}} \textbf{3}, 730 (2012).

\bibitem{oleary2021}
C.~M. O'Leary, G.~T. Martinez, E.~Liberti, M.~J. Humphry, A.~I. Kirkland, and
  P.~D. Nellist, \enquote{Contrast transfer and noise considerations in
  focused-probe electron ptychography,}
  {\protect\JournalTitle{Ultramicroscopy}} \textbf{221}, 113189 (2021).

\bibitem{shi2013}
Y.~Shi, T.~Li, Y.~Wang, Q.~Gao, S.~Zhang, and H.~Li, \enquote{Optical image
  encryption via ptychography,} {\protect\JournalTitle{Optics Letters}}
  \textbf{38}, 1425 (2013).

\bibitem{rawat2015}
N.~Rawat, I.-C. Hwang, Y.~Shi, and B.-G. Lee, \enquote{Optical image encryption
  via photon-counting imaging and compressive sensing based ptychography,}
  {\protect\JournalTitle{Journal of Optics}} \textbf{17}, 065704 (2015).

\bibitem{zhu2019}
Y.~Zhu, W.~Xu, and Y.~Shi, \enquote{High-capacity encryption system based on
  single-shot-ptychography encoding and {{QR}} code,}
  {\protect\JournalTitle{Optics Communications}} \textbf{435}, 426--432 (2019).

\bibitem{aidukas2019}
T.~Aidukas, P.~C. Konda, A.~R. Harvey, M.~J. Padgett, and P.-A. Moreau,
  \enquote{Phase and amplitude imaging with quantum correlations through
  {{Fourier Ptychography}},} {\protect\JournalTitle{Scientific Reports}}
  \textbf{9}, 10445 (2019).

\bibitem{fernandes2019}
M.~F. Fernandes and L.~Neves, \enquote{Ptychography of pure quantum states,}
  {\protect\JournalTitle{Scientific Reports}} \textbf{9}, 16066 (2019).

\bibitem{fernandes2020}
M.~F. Fernandes, M.~A. {Sol{\'i}s-Prosser}, and L.~Neves,
  \enquote{Ptychographic reconstruction of pure quantum states,}
  {\protect\JournalTitle{Optics Letters}} \textbf{45}, 6002 (2020).

\bibitem{wangya-li2013}
{Wang Ya-Li}, {Shi Yi-Shi}, {Li Tuo}, {Gao Qian-Kun}, {Xiao Jun}, and {Zhang
  San-Guo}, \enquote{Research on the key parameters of illuminating beam for
  imaging via ptychography in visible light band,} {\protect\JournalTitle{Acta
  Physica Sinica}} \textbf{62}, 064206 (2013).

\bibitem{huang2014}
X.~Huang, H.~Yan, R.~Harder, Y.~Hwu, I.~K. Robinson, and Y.~S. Chu,
  \enquote{Optimization of overlap uniformness for ptychography,}
  {\protect\JournalTitle{Optics Express}} \textbf{22}, 12634 (2014).

\bibitem{sidorenko2016}
P.~Sidorenko and O.~Cohen, \enquote{Single-shot ptychography,}
  {\protect\JournalTitle{Optica}} \textbf{3}, 9 (2016).

\bibitem{fienup1997}
J.~R. Fienup, \enquote{Invariant error metrics for image reconstruction,}
  {\protect\JournalTitle{Applied Optics}} \textbf{36}, 8352 (1997).

\end{thebibliography}

\end{document}